\documentclass{article}
\usepackage{amssymb}
\usepackage{amsmath}

\usepackage{epsf}
\usepackage{epsfig}
\usepackage{floatflt}
\usepackage{float}

\input{tcilatex}
\begin{document}

\begin{center}
\bigskip

{\huge Time-dependent polynomials with \textit{one double} root, and related new
solvable systems of nonlinear evolution equations}

\bigskip

\textbf{Oksana Bihun}$^{\left( a,1\right) }$, \textbf{Francesco Calogero}$%
^{\left( b,c,2,3\right) }$\bigskip

$^{a}~$Department of Mathematics, University of Colorado, Colorado Springs,

1420 Austin Bluffs Pkway, Colorado Springs, CO 80918, USA

$^{1}~$obihun@UCCS.edu

$^{b}~$Physics Department, University of Rome \textquotedblleft La
Sapienza\textquotedblright , Italy

$^{c}~$Istituto Nazionale di Fisica Nucleare, Sezione di Roma, Italy

$^{2}~$francesco.calogero@roma1.infn.it, $^{3}~$%
francesco.calogero@uniroma1.it

\bigskip

\textit{Abstract}
\end{center}

Recently \textit{new solvable} systems of nonlinear evolution
equations---including ODEs, PDEs and systems with discrete time---have been
introduced. These findings are based on certain convenient formulas
expressing the $k$-th time-derivative of a root of a time-dependent monic
polynomial in terms of the $k$-th time-derivative of the coefficients of the
same polynomial and of the roots of the same polynomial as well as their
time-derivatives of order less than $k$. These findings were restricted to
the case of \textit{generic} polynomials \textit{without any multiple} root.
In this paper some of these findings---those for $k=1$\ and $k=2$\textbf{---}%
are extended to polynomials featuring \textit{one double} root; and a few
representative examples are reported of \textit{new solvable} systems of
nonlinear evolution equations.



\bigskip

\section{Introduction}

Recently new classes of systems of \textit{solvable} evolution
equations---including ordinary differential equations (ODEs), partial
differential equations (PDEs) and systems evolving in \textit{discrete}
time---have been identified \cite{1}-\cite{14}. The basic idea of these
recent developments is quite simple and rather old. Consider a
time-dependent monic polynomial of degree $N$ in the \textit{complex}
variable $z$, which is then characterized by $N$ time-dependent coefficients 
$y_{m}\left( t\right) $ and $N$ zeros $x_{n}\left( t\right) .$ [Here and
hereafter $N$ is a fixed positive integer larger than unity ($N\geq 2$),
indices such as $m$ and $n$ range over the integers from $1$ to $N$ unless
otherwise indicated, the \textit{real} variable $t$ is ``time'', superimposed
dots indicate time-differentiations, and all other quantities are generally 
\textit{complex} numbers]. Assume then that the time evolution of the $N$ 
\textit{coefficients} $y_{m}\left( t\right) $ of this polynomial evolve in
time according to equations of motion which are in some sense
``solvable''---for instance by algebraic operations, or via some other
appropriate and convenient technique. It is then often the case that these
evolutions are, in some sense, ``simple and interesting'': for instance, 
\textit{Hamiltonian} and possibly \textit{integrable} and/or \textit{multiply periodic}, 
\textit{completely periodic} or even \textit{isochronous}. Look then at the
corresponding evolution of the $N$ \textit{zeros} $x_{n}\left( t\right) $.
It is generally more complicated (``more nonlinear''), yet it generally inherits the
properties of the evolution of the \textit{coefficients }$y_{m}\left(
t\right) $: hence it is, in some sense, also ``simple and interesting'',
therefore worth of identification and further study. This approach was
introduced 4 decades ago \cite{15}-\cite{17} to identify \textit{%
integrable/solvable} many-body problems characterized by evolution equations
of Newtonian type---``accelerations equal forces''---describing $N$ points
moving in the \textit{complex} plane (or equivalently in the \textit{real}
plane) identified with the $N$ \textit{zeros} $x_{n}\left( t\right) $ of
time-dependent monic polynomials the $N$ \textit{coefficients} $y_{m}\left(
t\right) $ of which evolve in time according to a system of \textit{linear}
ODEs. At the time this restriction to a \textit{linear} evolution of the 
\textit{coefficients} was essential in order to be able to write \textit{%
explicitly} the corresponding evolution of the \textit{zeros}. Only recently
a simple technique has been introduced \cite{1} \cite{7}, which allows to
obtain in \textit{explicit} form the equations of motion of the \textit{zeros%
} of a time-dependent monic polynomial the \textit{coefficients} of which
evolve according to \textit{nonlinear} equations of motion. This opened the
way to the identification and investigation of large classes of \textit{new
solvable nonlinear} evolution equations, as mentioned above \cite{1}-\cite%
{14}.

This development was however restricted so far to the consideration of 
\textit{generic} time-dependent polynomials, the \textit{zeros} of which are
\textit{all different among themselves}, except possibly at some special times
corresponding to collisions of some of the moving \textit{zeros}.

In the present paper a first step is made towards the elimination of this
restriction, by considering polynomials which feature \textit{one double}
zero, opening thereby the possibility to identify additional classes of
nonlinear evolution amenable to exact treatments; and some such examples are
reported.

This generalization is described in the following \textbf{Section 2}, and
some examples of \textit{new }systems of \textit{solvable nonlinear }%
evolution equations are identified and investigated in \textbf{Section 3}.
The last \textbf{Section 4} (``Outlook'') provides a terse overview of further
developments.

\bigskip

\section{Monic time-dependent polynomials with \textit{one double} root, and related
formulas}

In this \textbf{Section 2} we report and prove formulas---of a type which is
particularly convenient for the identification and investigation of \textit{%
new solvable} evolution equations \cite{1}-\cite{14}---which relate the time
evolution of the \textit{zeros} of a (\textit{nongeneric}) monic polynomial
featuring---for all time---\textit{one double} \textit{zero}, to the
time-evolution of its \textit{coefficients}. This \textbf{Section 2} is
divided into 2 Subsections, in which we treat this problem in order of
increasing complexity: in \textbf{Subsection 2.1,} the  simplest problem
of a time-dependent polynomial of \textit{third} degree with a \textit{%
double zero}; and in \textbf{Subsection 2.2}, the case of a time-dependent
polynomial of \textit{arbitrary} degree $N+1$ with \textit{one double zero}.
The extension of these findings to the most general case---a time-dependent
polynomial of \textit{arbitrary} degree featuring \textit{several} \textit{%
zeros of arbitrary multiplicities}---is a nontrivial undertaking: this task
shall eventually be treated---by ourselves or by others---in subsequent
publications.

\bigskip

\subsection{The monic time-dependent polynomial of \textit{third} degree
with a \textit{double} zero, and related formulas}

It is convenient to start from the  simplest case, that of a monic
polynomial of degree $3$ featuring a \textit{double} zero: 
\begin{subequations}
\label{p3}
\begin{equation}
p_{3}\left( z;t\right) =z^{3}+y_{1}\left( t\right) ~z^{2}+y_{2}\left(
t\right) ~z+y_{3}\left( t\right) ~,  \label{p3y}
\end{equation}%
\begin{equation}
p_{3}\left( z;t\right) =\left[ z-x_{1}\left( t\right) \right] ^{2}~\left[
z-x_{2}\left( t\right) \right] ~.  \label{p3x}
\end{equation}%
Indeed, this case is simple enough to write the relevant formulas without
explanations, yet it is sufficient to highlight the  complications
that make the case with \textit{multiple} zeros rather different from the
case of \textit{generic} polynomials---hence, featuring \textit{no} multiple
zeros---previously treated \cite{1}-\cite{14}. 

Hereafter the \textit{explicit%
} indication of the time-dependence of the various quantities will be
omitted whenever we feel that this omission---even if, occasionally, applied 
\textit{inconsistently} within the same formula---is unlikely to cause
misunderstandings.

The 3 coefficients $y_{1},~y_{2},~y_{3}$ are of course given by the
following formulas in terms of the two zeros $x_{1}$ and $x_{2}$: 
\end{subequations}
\begin{subequations}
\label{yydot123}
\begin{equation}
y_{1}=-\left( 2~x_{1}+x_{2}\right) ~,~~y_{2}=\left( x_{1}\right)
^{2}+2~x_{1}~x_{2}~,~~y_{3}=-\left( x_{1}\right) ^{2}~x_{2}~;  \label{y123}
\end{equation}%
and these equations imply  the following expressions of the time
derivatives $\dot{y}_{m}$:%
\begin{eqnarray}
\dot{y}_{1} &=&-\left( 2~\dot{x}_{1}+\dot{x}_{2}\right) ~,~~\hspace{0in}~%
\dot{y}_{2}=2~\left[ \dot{x}_{1}~\left( x_{1}+x_{2}\right) +\dot{x}_{2}~x_{1}%
\right] ~,  \notag \\
\dot{y}_{3} &=&-\left[ 2~\dot{x}_{1}~x_{1}~x_{2}+\dot{x}_{2}~\left(
x_{1}\right) ^{2}\right] ~.  \label{ydot123}
\end{eqnarray}

There hold moreover the following formulas: 
\end{subequations}
\begin{equation}
p_{3}\left( x_{n};t\right) =\left( x_{n}\right) ^{3}+y_{1}~\left(
x_{n}\right) ^{2}+y_{2}~x_{n}+y_{3}=0~,~~~n=1,2~;  \label{xn}
\end{equation}%
\begin{eqnarray}
&&p_{3,z}\left( z;t\right) =3~z^{2}+2~y_{1}~z+y_{2}  \notag \\
&=&2~\left( z-x_{1}\right) ~\left( z-x_{2}\right) +\left( z-x_{1}\right)
^{2}~,  \label{p3z}
\end{eqnarray}%
\begin{eqnarray}
&&p_{3,t}\left( z;t\right) =\dot{y}_{1}~z^{2}+\dot{y}_{2}~z+\dot{y}_{3} 
\notag \\
&=&-2~\dot{x}_{1}~\left( z-x_{1}\right) ~\left( z-x_{2}\right) -\dot{x}%
_{2}~\left( z-x_{1}\right) ^{2}~.  \label{p3t}
\end{eqnarray}%
Above and hereafter appended variables preceded by a comma denote 
(partial) differentiations with respect to that variable.

For $z=x_{1}$ respectively for $z=x_{2}$ the last $2$ formulas yield the
following relations: 
\begin{subequations}
\begin{equation}
3~\left( x_{1}\right) ^{2}+2~y_{1}~x_{1}+y_{2}=0~,  \label{Eqx1}
\end{equation}%
\begin{equation}
3~\left( x_{2}\right) ^{2}+2~y_{1}~x_{2}+y_{2}=\left( x_{1}-x_{2}\right)
^{2}~;
\end{equation}%
\end{subequations}
\begin{subequations}
\begin{equation}
\dot{y}_{1}~\left( x_{1}\right) ^{2}+\dot{y}_{2}~x_{1}+\dot{y}_{3}=0~,
\end{equation}%
\begin{equation}
\dot{y}_{1}~\left( x_{2}\right) ^{2}+\dot{y}_{2}~x_{2}+\dot{y}_{3}=-\dot{x}%
_{2}~\left( x_{1}-x_{2}\right) ^{2}~.
\end{equation}

It is then a matter of trivial algebra to obtain the following $3$
(different!) systems of evolution equations, which relate the time evolution
of the $2$ zeros $x_{1}\left( t\right) $ and $x_{2}\left( t\right) $ to the
evolution of $2$ out of the $3$ coefficients $y_{1}\left( t\right)
,~y_{2}\left( t\right) ,~y_{3}\left( t\right) $: 
\end{subequations}
\begin{equation}
\dot{x}_{1}=-\frac{2~x_{1}~\dot{y}_{1}+\dot{y}_{2}}{2~\left(
x_{1}-x_{2}\right) }~,~~~\dot{x}_{2}=\frac{\left( x_{1}+x_{2}\right) ~\dot{y}%
_{1}+\dot{y}_{2}}{x_{1}-x_{2}}~;  \label{EqMot11}
\end{equation}%
\begin{equation}
\dot{x}_{1}=-\frac{\left( x_{1}\right) ^{2}~\dot{y}_{1}-\dot{y}_{3}}{%
2~x_{1}~\left( x_{1}-x_{2}\right) }~,~~~\dot{x}_{2}=\frac{x_{1}~x_{2}~\dot{y}%
_{1}-\dot{y}_{3}}{x_{1}~\left( x_{1}-x_{2}\right) }~;  \label{EqMot12}
\end{equation}%
\begin{equation}
\dot{x}_{1}=\frac{x_{1}~\dot{y}_{2}+2~\dot{y}_{3}}{2~x_{1}~\left(
x_{1}-x_{2}\right) }~,~~~\dot{x}_{2}=-\frac{x_{1}~x_{2}~\dot{y}_{2}+\left(
x_{1}+x_{2}\right) ~\dot{y}_{3}}{\left( x_{1}\right) ^{2}~\left(
x_{1}-x_{2}\right) }~.  \label{EqMot13}
\end{equation}

Analogous equations can be obtained for higher-order time-derivatives; the
relevant expressions become progressively more complicated as the order of
differentiation increases. Here we report the equations for the second
time-derivatives, in view of their relevance in the study of many-body
dynamics due to their relations to the Newtonian equations of motion of
classical mechanics (``accelerations equal forces''). They read as follows: 
\begin{eqnarray}
&&\ddot{x}_{1}=\frac{2~\dot{x}_{1}~\left( \dot{x}_{1}+2~\dot{x}_{2}\right)
-2~x_{1}~\ddot{y}_{1}-\ddot{y}_{2}}{2~\left( x_{1}-x_{2}\right) }~,\notag\\
&&\ddot{x}_{2}=\frac{-2~\dot{x}_{1}~\left( \dot{x}_{1}+2~\dot{x}_{2}\right)
+\left( x_{1}+x_{2}\right) ~\ddot{y}_{1}+\ddot{y}_{2}}{x_{1}-x_{2}}~;
\label{xdotdot1}
\end{eqnarray}%
\begin{eqnarray}
&&\ddot{x}_{1}=\frac{2~\dot{x}_{1}~\left( \dot{x}_{1}~x_{2}+2~\dot{x}%
_{2}~x_{1}\right) -\left( x_{1}\right) ^{2}~\ddot{y}_{1}+\ddot{y}_{3}}{%
2~x_{1}~\left( x_{1}-x_{2}\right) }~, \notag\\
&&\ddot{x}_{2}=\frac{-2~\dot{x}_{1}~\left( \dot{x}_{1}~x_{2}+2~\dot{x}%
_{2}~x_{1}\right) +x_{1}~x_{2}~\ddot{y}_{1}-\ddot{y}_{3}}{x_{1}~\left(
x_{1}-x_{2}\right) }~; \label{xdotdot2}
\end{eqnarray}
\begin{eqnarray}
&&\ddot{x}_{1}=\frac{-2~x_{1}~\dot{x}_{1}~\left( \dot{x}_{1}-2~\dot{x}%
_{2}\right) +4~\left( \dot{x}_{1}\right) ^{2}~x_{2}+x_{1}~\ddot{y}_{2}+2~%
\ddot{y}_{3}}{2~x_{1}~\left( x_{1}-x_{2}\right) }~,  \notag\\
&&\ddot{x}_{2}=-\frac{2~\dot{x}_{1}~\left[ \dot{x}_{1}~\left( x_{2}\right)
^{2}+2~\dot{x}_{2}~\left( x_{1}~\right) ^{2}\right] +x_{1}~x_{2}~\ddot{y}%
_{2}+\left( x_{1}+x_{2}\right) ~\ddot{y}_{3}}{\left( x_{1}\right)
^{2}~\left( x_{1}-x_{2}\right) }~.  \label{xdotdot3}
\end{eqnarray}
These $3$ (different!) systems of $2$ coupled ODEs provide the tools to
identify and investigate \textit{new} systems of $2$ equations of motions of
potential theoretical or applicative interest (although they have been
mainly discussed here to introduce the treatment of analogous---but more
general---\textit{new} systems of $N$ nonlinear evolution equations). The
idea---as in \cite{1}-\cite{17}---is to assume that $2$ of the $3$
quantities $y_{m}\left( t\right) $ evolve in time according to a system of
evolution equations amenable to exact treatments; say 
\begin{equation}
\ddot{y}_{1}=f_{1}\left( \dot{y}_{1},~\dot{y}_{2};~y_{1},~y_{2}\right) ~,~~~%
\ddot{y}_{2}=f_{2}\left( \dot{y}_{1},~\dot{y}_{2};~y_{1},~y_{2}\right) ~,
\label{SolvNewty1}
\end{equation}%
with the $2$ functions $f_{1}\left( \dot{y}_{1},~\dot{y}_{2};~y_{1},~y_{2}%
\right) $ and $f_{2}\left( \dot{y}_{1},~\dot{y}_{2};~y_{1},~y_{2}\right) $
such that this system can be explicitly solved (see examples in Section 3).
Then, by inserting these expressions in the right hand side of system (\ref%
{xdotdot1}),
one obtains the following system of $2$, generally highly
nonlinear, evolution equations: 
\begin{eqnarray}
\ddot{x}_{1}&=&\frac{1}{2}~\left(
x_{1}-x_{2}\right)^{-1} \Big[
2~\dot{x}_{1}~\left( \dot{x}_{1}+2~\dot{x}_{2}\right)
-2~x_{1}~f_{1}\left( \dot{y}_{1},~\dot{y}_{2};~y_{1},~y_{2}\right)
\notag\\
&&-f_{2}\left( \dot{y}_{1},~\dot{y}_{2};~y_{1},~y_{2}\right) \Big]~, \notag\\
\ddot{x}_{2}&=&(x_{1}-x_{2})^{-1} \Big[
-2~\dot{x}_{1}~\left( \dot{x}_{1}+2~\dot{x}_{2}\right)
+\left( x_{1}+x_{2}\right) ~f_{1}\left( \dot{y}_{1},~\dot{y}%
_{2};~y_{1},~y_{2}\right) \notag\\
&&+f_{2}\left( \dot{y}_{1},~\dot{y}%
_{2};~y_{1},~y_{2}\right) \Big]~
\label{Solv2BodyProb1}
\end{eqnarray}%
with, in the right-hand sides, the quantities $\dot{y}_{1},~\dot{y}%
_{2},~y_{1},~y_{2}$ replaced by their explicit expressions (see (\ref%
{ydot123}) and (\ref{y123})) in terms of $x_{1}$ and $x_{2}$ and their time
derivatives. This is then one of the $3$ \textit{new solvable} systems of $2$
nonlinearly coupled second-order (``Newtonian'') equations of motion satisfied
by the quantities $x_{1}\left( t\right) $ and $x_{2}\left( t\right) $ (the
other two such systems obtain of course in an analogous manner from (\ref%
{xdotdot2}) and (\ref{xdotdot3}) rather than (\ref{xdotdot1}); see below).
Let us now explain how the solution of this $2$-body problem, (\ref%
{Solv2BodyProb1}), can be achieved.

\textit{Step (i)}. Given the initial values $x_{n}\left( 0\right) $ of the $2$ zeros $%
x_{n}\left( t\right) $ as well as the initial values of their velocities,
via the formulas (\ref{y123}) and (\ref{ydot123}) the initial values $%
y_{m}\left( 0\right) $ and $\dot{y}_{m}\left( 0\right) $ with $m=1,2,3$ are
(easily) computed.

\textit{Step (ii)}. From the initial values $y_{m}\left( 0\right) $ and $\dot{y}%
_{m}\left( 0\right) $ with $m=1,2$ the values of $y_{m}\left( t\right) $%
---and then as well of $\dot{y}_{m}\left( t\right) $---are computed by
solving the---assumedly \textit{solvable}---system of evolution equations (%
\ref{SolvNewty1}).

\textit{Step (iii)}. From the knowledge of $y_{1}\left( t\right) $ and $y_{2}\left(
t\right) $ the value of $x_{1}\left( t\right) $ is computed by solving the 
\textit{quadratic} equation (\ref{Eqx1}). There obtain, for every value of $%
t,$ \textit{two} different values of $x_{1}\left( t\right) ;$ and by
following them, by continuity in $t,$ all the way back to $t=0$---and by
comparing the value $x_{1}\left( 0\right) $ yielded by this procedure with
the initial datum $x_{1}\left( 0\right) $---the actual (continuous) solution 
$x_{1}\left( t\right) $ is identified.

\textit{Step (iv)}. The solution $x_{2}\left( t\right) $ for all time is then
immediately obtained, for instance, from the known functions $x_{1}\left(
t\right) $ and $y_{1}\left( t\right) ,$ via the \textit{first} of the $3$
equations (\ref{y123}).

The way to solve the other two systems, (\ref{xdotdot2}) and (\ref{xdotdot3}%
), is analogous but not quite identical, so let us tersely detail how the
method works for the \textit{second} system (\ref{xdotdot2}).

So, we now take as point of departure the system 
\begin{equation}
\ddot{y}_{1}=f_{1}\left( \dot{y}_{1},~\dot{y}_{3};~y_{1},~y_{3}\right) ~,~~~%
\ddot{y}_{3}=f_{3}\left( \dot{y}_{1},~\dot{y}_{3};~y_{1},~y_{3}\right) ~,
\label{SolvNewty2}
\end{equation}%
with the $2$ functions $f_{1}\left( \dot{y}_{1},~\dot{y}_{3};~y_{1},~y_{3}%
\right) $ and $f_{3}\left( \dot{y}_{1},~\dot{y}_{3};~y_{1},~y_{3}\right) $
such that this system can be explicitly solved (see examples below). Then,
by inserting these expressions in the right hand side of the system (\ref%
{xdotdot1}), one obtains the following system of $2$, generally highly
nonlinear, evolution equations: 
\begin{eqnarray}
\ddot{x}_{1}&=&\left[2 x_{1}\left( x_{1}-x_{2}\right)\right]^{-1}
\Big[2~\dot{x}_{1}~\left( \dot{x}_{1}~x_{2}+2~\dot{x}%
_{2}~x_{1}\right) -\left( x_{1}\right) ^{2}~f_{1}\left( \dot{y}_{1},~\dot{y}%
_{3};~y_{1},~y_{3}\right) \notag\\
&&+f_{3}\left( \dot{y}_{1},~\dot{y}%
_{3};~y_{1},~y_{3}\right) \Big]~, \notag\\
\ddot{x}_{2}&=&\left[ x_{1}\left( x_{1}-x_{2}\right)\right]^{-1}
\Big[-2~\dot{x}_{1}~\left( \dot{x}_{1}~x_{2}+2~\dot{x}%
_{2}~x_{1}\right) +x_{1}~x_{2}~f_{1}\left( \dot{y}_{1},~\dot{y}%
_{3};~y_{1},~y_{3}\right) \notag\\
&&-f_{3}\left( \dot{y}_{1},~\dot{y}%
_{3};~y_{1},~y_{3}\right) \Big]~;
\label{Solv2BodyProb2}
\end{eqnarray}%
with, in the right-hand sides, the quantities $\dot{y}_{1},~\dot{y}%
_{3},~y_{1},~y_{3}$ replaced by their explicit expressions (see (\ref%
{ydot123}) and (\ref{y123})) in terms of $x_{1}$ and $x_{2}$ and their time
derivatives. This is then the \textit{second} one of the $3$ \textit{new
solvable} systems of $2$ nonlinearly coupled second-order (``Newtonian'')
equations of motion satisfied by the quantities $x_{1}\left( t\right) $ and $%
x_{2}\left( t\right) $.

Let us now explain how the solution of this $2$-body problem, (\ref%
{Solv2BodyProb2}), can be achieved.

\textit{Step (i)}. Given the initial values $x_{n}\left( 0\right) $ of the $2$ zeros $%
x_{n}\left( t\right) $ as well as the initial values of their velocities,
via the formulas (\ref{y123}) and (\ref{ydot123}) the initial values $%
y_{m}\left( 0\right) $ and $\dot{y}_{m}\left( 0\right) $ with $m=1,2,3$ are
(easily) computed.

\textit{Step (ii)}. From the initial values $y_{m}\left( 0\right) $ and $\dot{y}%
_{m}\left( 0\right) $ with $m=1,3$ the values of $y_{m}\left( t\right) $%
---and then as well of $\dot{y}_{m}\left( t\right) $---are computed by
solving the---assumedly solvable---system of evolution equations (\ref%
{SolvNewty2}).

\textit{Step (iii)}. From the knowledge of $y_{1}\left( t\right) $ and $y_{3}\left(
t\right) $ the value of $x_{1}\left( t\right) $ is computed as the root of
the \textit{cubic} equation 
\begin{equation}
2~\left( x_{1}\right) ^{3}+y_{1}~\left( x_{1}\right) ^{2}-y_{3}=0~,
\end{equation}%
which is implied by (\ref{Eqx1}) via the \textit{first} of the $3$ equations
(\ref{y123}). By solving this \textit{cubic} equation there obtain, for
every value of $t,$ three different values of $x_{1}\left( t\right) ;$ and
by following them, by continuity in $t,$ all the way back to $t=0$---and by
then comparing the value $x_{1}\left( 0\right) $ yielded by this procedure
with the initial datum $x_{1}\left( 0\right) $---the actual (continuous)
solution $x_{1}\left( t\right) $ is identified.

\textit{Step (iv)}. The solution $x_{2}\left( t\right) $ for all time is then
immediately obtained, for instance, from the known functions $x_{1}\left(
t\right) $ and $y_{1}\left( t\right) ,$ via the first of the $3$ equations (%
\ref{y123}).

The treatment of the \textit{third} of the $3$ solvable systems the
equations of motion of which read 
\begin{eqnarray}
&&\ddot{x}_{1}=\left[ 2~x_{1}~\left( x_{1}-x_{2}\right) \right] ^{-1}~\Big[
-2~x_{1}~\dot{x}_{1}~\left( \dot{x}_{1}-2~\dot{x}_{2}\right) +4~\left( \dot{x%
}_{1}\right) ^{2}~x_{2}  \notag \\
&& +x_{1}~f_{2}\left( \dot{y}_{2},~\dot{y}_{3};~y_{2},~y_{3}\right)
+2~f_{3}\left( \dot{y}_{2},~\dot{y}_{3};~y_{2},~y_{3}\right) \Big] ~, \notag\\
&&\ddot{x}_{2}=-\left[ \left( x_{1}\right) ^{2}~\left( x_{1}-x_{2}\right) %
\right] ^{-1}~\Big\{ 2~\dot{x}_{1}~\left[ \dot{x}_{1}~\left( x_{2}\right)
^{2}+2~\dot{x}_{2}~\left( x_{1}~\right) ^{2}\right]  \notag \\
&& +x_{1}~x_{2}~f_{2}\left( \dot{y}_{2},~\dot{y}_{3};~y_{2},~y_{3}%
\right) +\left( x_{1}+x_{2}\right) ~f_{3}\left( \dot{y}_{2},~\dot{y}%
_{3};~y_{2},~y_{3}\right) \Big\} ~,
\label{Solv2BodyProb3}
\end{eqnarray}%
---corresponding to the equations of motion 
\begin{equation}
\ddot{y}_{2}=f_{2}\left( \dot{y}_{2},~\dot{y}_{3};~y_{2},~y_{3}\right) ~,~~~%
\ddot{y}_{3}=f_{3}\left( \dot{y}_{2},~\dot{y}_{3};~y_{2},~y_{3}\right)
\label{SolvNewty3}
\end{equation}%
---is quite analogous, except for \textit{Step (iii)}, that now requires solving the
following \textit{cubic} equation to determine $x_{1}\left( t\right) $ in
terms of $y_{2}\left( t\right) $ and $y_{3}\left( t\right) $: 
\begin{equation}
\left( x_{1}\right) ^{3}-x_{1}~y_{2}-2~y_{3}=0~.
\end{equation}

It is plain from this treatment---see, if need be, the detailed discussions
of this question in \cite{1}-\cite{17}---that the $3$ two-body problems (\ref%
{Solv2BodyProb1}), (\ref{Solv2BodyProb2}) respectively (\ref{Solv2BodyProb3}%
)\ ``inherit'' all the nice properties---such as the property to be \textit{%
Hamiltonian}, the property of \textit{integrability} in the Hamiltonian
context, and various \textit{periodicity} properties including \textit{%
isochrony} and \textit{asymptotic isochrony}---possibly possessed by the
solvable systems (\ref{SolvNewty1}), (\ref{SolvNewty2}) respectively (\ref%
{SolvNewty3}).  Some specific examples are reported in \textbf{Subsection 3.1}

\bigskip

\subsection{The monic time-dependent polynomial of degree $N+1$ with a 
\textit{double} zero, and related formulas}

The monic time-dependent polynomial of degree $N+1$ featuring $N+1$ \textit{%
coefficients} and $N$ \textit{zeros}, one of which is \textit{double}, reads
as follows: 
\begin{subequations}
\label{pN+1zt}
\begin{equation}
p_{N+1}\left( z;t\right) =z^{N+1}+\sum_{m=1}^{N+1}\left[ y_{m}\left(
t\right) ~z^{N+1-m}\right] ~,  \label{pN+1zta}
\end{equation}%
\begin{equation}
p_{N+1}\left( z;t\right) =\left[ z-x_{1}\left( t\right) \right]
^{2}~\dprod\limits_{n=2}^{N}\left[ z-x_{n}\left( t\right) \right] ~.
\label{pN+1ztb}
\end{equation}%
Note that it features $N+1$ coefficients $y_{m}\left( t\right) $ (with $%
m=1,2,...,N+1$, see (\ref{pN+1zta})) and $N$ zeros $x_{n}\left( t\right) $
(with $n=1,2,...,N$, see (\ref{pN+1ztb})); all but one of these zeros have
unit multiplicity, the exceptional one having multiplicity $2$ and being
identified as $x_{1}\left( t\right) $. This of course implies that the $N+1$
coefficients $y_{m}\left( t\right) $ are, for all time $t,$ related to each
other by one constraint (see below). 
As in the previous section, we will occasionally omit the explicit indication of the $t$-dependence of $x_n=x_n(t)$ and $y_m=y_m(t).$

These formulas, (\ref{pN+1zt}), imply of course that the $N+1$ coefficients $%
y_{m}$ are expressed in terms of the $N$ zeros $x_{n}$ by the standard
formulas 
\end{subequations}
\begin{subequations}
\label{ymxSect22}
\begin{equation}
y_{m}=\left( -1\right) ^{m}~\sigma _{m}\left( \tilde{x}\right) ~,
\label{ymx}
\end{equation}%
where $\sigma _{m}\left( \tilde{x}\right) $ is the standard symmetric polynomial of 
degree $m$ in $N+1$ variables, evaluated at any permutation $\tilde{x}$ of the vector  $\left(
x_{1},~x_{1},~x_{2},...,~x_{N}\right) $ of the $N+1$ zeros
of $p_{N+1}\left( z;t\right) $ with $x_{1}$ repeated twice. For
instance, the first and last of these coefficients are expressed in terms of
the $N$ zeros $x_{n}$ as follows:%
\begin{equation}
y_{1}=-\left( 2~x_{1}+x_{2}+x_{3}+...+x_{N}\right) ~,  \label{y1}
\end{equation}%
\begin{equation}
y_{N+1}=\left( -\right) ^{N+1}\left( x_{1}\right) ^{2}~x_{2}~x_{3}...x_{N}~.
\label{yN+1}
\end{equation}%
Conversely---once the polynomial $p_{N+1}\left( z\right) $ has been assigned
via its $N+1$ coefficients $y_{m}$---then its $N$ zeros $x_{n}$ are uniquely
determined (up to permutations of the $N-1$ zeros $x_{2},~x_{3},~...,~x_{N}$%
). They can therefore be obtained from the $N+1$ coefficients $y_{m}$ by 
\textit{algebraic} operations, which however can generally be \textit{%
explicitly} performed only for small values of $N$. But a more relevant
issue for our purposes---which is treated at the end of this \textbf{Subsection
2.2}---is to show how the $N$ zeros $x_{n}$ with $n=1,2,...,N$ of the
polynomial (\ref{pN+1zt}), as well as one of the $N$ coefficients $y_{m}$%
---say, the coefficient $y_{\bar{m}}$, with $\bar{m}$ an arbitrarily assigned integer in
the range from $1$ to $N+1$---can be obtained by algebraic operations from
the $N$ coefficients $y_{m}$ with $m=1,2,...,\bar{m}-1,\bar{m}+1,...,N,N+1$
of this polynomial (\ref{pN+1zt}) (taking of course advantage of the fact
that this polynomial features a \textit{double} zero). Note the special role (\textit{not}!)
played by the coefficient $y_{\bar{m}}$, with $\bar{m}$ an \textit{arbitrarily assigned}
integer in the range from $1$ to $N+1.$ Our task is to derive convenient
expressions of the first respectively the second time derivative, $\dot{x}%
_{n}\left( t\right) $ respectively $\ddot{x}_{n}\left( t\right) $ (for every
given $n$ in the range from $1$ to $N$), in terms of the first respectively
the second derivatives, $\dot{y}_{m}\left( t\right) $ respectively $\ddot{y}%
_{m}\left( t\right) ,$ of the $N$ coefficients $y_{m}\left( t\right) $ with $%
m=1,2,...,\bar{m}-1,\bar{m}+1,...,N+1$ (hence of all the $N+1$ coefficients $%
y_{m}\left( t\right) $ with the exclusion of the special coefficient $y_{%
\bar{m}}\left( t\right) $). At the end of this section we also discuss how
the $N$ zeros $x_{n}$ with $n=1,2,...,N$ of the polynomial (\ref{pN+1zt}),
as well as the coefficient $y_{\bar{m}}$ of this polynomial, can be obtained
via \textit{algebraic} operations from the $N$ coefficients $y_{m}$ of this
polynomial (\ref{pN+1zt}) with $m=1,2,...\bar{m}-1,~\bar{m}+1,...N+1$
(taking of course advantage of the fact that this polynomial features a 
\textit{double} zero).

The $t$-derivatives of the two versions, (\ref{pN+1zta}) and (\ref{pN+1ztb})
of (\ref{pN+1zt}), read of course as follows: 
\end{subequations}
\begin{subequations}
\begin{equation}
p_{N+1,t}\left( z;t\right) =\sum_{m=1}^{N+1}\left[ \dot{y}_{m} ~z^{N+1-m}\right] ~,
\end{equation}%
\begin{eqnarray}
&&p_{N+1,t}\left( z;t\right) =-2~\dot{x}_{1} ~\left(
z-x_{1}  \right) ~\dprod\limits_{n=2}^{N}\left( z-x_{n}  \right) \notag \\
&&-\left( z-x_{1}  \right) ^{2}~\sum_{n=2}^{N}\left[ \dot{x}%
_{n}  ~\dprod\limits_{\ell =2,~\ell \neq n}^{N}\left(
z-x_{\ell }  \right) \right] ~.
\end{eqnarray}%
Hence, equating these two formulas for $z=x_{1}  ,$
respectively for $z=x_{n}  ,$ we obtain the following two
identities: 
\end{subequations}
\begin{subequations}
\begin{equation}
\sum_{m=1}^{N+1}\left[ \dot{y}_{m}  ~\left( x_{1}  \right) ^{N+1-m}\right] =0~,
\end{equation}%
\begin{eqnarray}
&&\sum_{m=1}^{N+1}\left[ \dot{y}_{m}  ~\left( x_{n}  \right) ^{N+1-m}\right]   \notag \\
&&=-\left( x_{n}  -x_{1}  \right) ^{2}~\left[
\dot{x}_{n}  ~\dprod\limits_{\ell =2,~\ell \neq n}^{N}\left(
x_{n}  -x_{\ell }  \right) \right],\notag\\
&& n=2,...,N~. 
\end{eqnarray}%
Then, by subtracting the first of these two formulas multiplied by $\left(
x_{1}  \right) ^{-\left( N+1-\bar{m}\right) }$ from the second
multiplied by $\left( x_{n}  \right) ^{-\left( N+1-\bar{m}%
\right) }$ we obtain the identity 
\end{subequations}
\begin{eqnarray}
&&-\left( x_{n}  -x_{1}  \right) ^{2}~\left[ 
\dot{x}_{n}  ~\dprod\limits_{\ell =2,~\ell \neq n}^{N}\left(
x_{n}  -x_{\ell }  \right) \right] ~\left(
x_{n}  \right) ^{-\left( N+1-\bar{m}\right) }  \notag \\
&=&\sum_{m=1}^{N+1}\left\{ \dot{y}_{m}  ~\left[ \left(
x_{n}  \right) ^{\bar{m}-m}-\left( x_{1}  \right)
^{\bar{m}-m}\right] \right\} ~,~~~n=2,...,N~,
\label{tfutd}
\end{eqnarray}%
hence the following expression of the derivatives $\dot{x}_{n}\left(
t\right) $ with $n=2,...,N:$%
\begin{subequations}
\begin{eqnarray}
&&\dot{x}_{n}  =- \left[ \dprod\limits_{\ell =1,~\ell
\neq n}^{N}\left( x_{n}  -x_{\ell }  \right)^{-1} %
\right] ~\left( x_{n}  \right) ^{N+1-\bar{m}%
}\cdot  \notag \\
&&\cdot \sum_{m=1,~m\neq \bar{m}}^{N+1}\left\{ \dot{y}_{m} 
~\left[ \frac{\left( x_{n}  \right) ^{\bar{m}-m}-\left(
x_{1}  \right) ^{\bar{m}-m}}{x_{n}  -x_{1}  }\right] \right\} ~,  \notag \\
n &=&2,...,N~,  \label{xndota}
\end{eqnarray}%
or, equivalently,%
\begin{eqnarray}
&&\dot{x}_n=-\left[\prod_{\ell=1, \ell \neq n}^N (x_n-x_\ell)^{-1} \right] (x_n)^{N+1-\bar{m}} \notag\\
&&\Bigg\{
\sum_{m=1}^{\bar{m}-1}\left[ \dot{y}_m \sum_{j=0}^{\bar{m}-m-1} (x_n)^{\bar{m}-m-1-j} (x_1)^j \right] \notag\\
&&-\sum_{m=\bar{m}+1}^{N+1}\left[ \dot{y}_m \sum_{j=0}^{m-\bar{m}-1}(x_n)^{\bar{m}-m+j} (x_1)^{-(j+1)}\right]
\Bigg\}.\notag\\
&&n=2,\ldots,N.
 \label{xndotb}
\end{eqnarray}
 \label{xndot}
\end{subequations}
This is the first key formula that expresses the first $t$-derivative $\dot{x%
}_{n}\left( t\right)  $ of the $N-1$ zeros $x_{n} \left( t\right) $ (with $%
n=2,...,N$) in terms of the first $t$-derivatives of the $N~$coefficients $%
y_{m}\left( t\right) $ (with $m\neq \bar{m})$: note that we evidenced the
important fact that the quantity $\dot{y}_{m}\left( t\right) $ with $m=\bar{m%
}$ does \textit{not} enter in these equations, since for $m=\bar{m}$ the summand in
the right-hand side of (\ref{xndota}) clearly \textit{vanishes} (and the second sum
in the right-hand side of (\ref{xndotb}) likewise vanishes since it is
empty).
\vspace{2mm}

\textbf{Remark 2.1.1}. Above and hereafter we assume for simplicity that $%
x_{1}\left( t\right) $ \textit{never} vanishes. Since $x_{1}\left( t\right) $
is by definition the \textit{double} zero of the polynomial (20), clearly a
sufficient condition to guarantee this is to restrict attention to time
evolutions of the two ``highest'' coefficients of this polynomial, $%
y_{N+1}\left( t\right) $ and $y_{N}\left( t\right) ,$ such that they \textit{%
never} vanish \textit{simultaneously}. $\blacksquare $

Next, let us derive an analogous formula for $\dot{x}_{1}\left( t\right) .$
To this end---and also for future developments---we now report the following
formulas expressing the first and second $t$-derivative of the first $z$%
-derivative of the rational functions $z^{\bar{m}-N-1}~p_{N+1}\left(
z;t\right) $ (see (\ref{pN+1zta})), which clearly read as follows: 
\begin{subequations}
\begin{equation}
\frac{\partial ^{2}}{\partial t~\partial z}\left[ z^{\bar{m}%
-N-1}~p_{N+1}\left( z;t\right) \right] =\sum_{m=1,~m\neq \bar{m}}^{N+1}\left[
\left( \bar{m}-m\right) ~\dot{y}_{m}  ~z^{\bar{m}-m-1}\right]
~,  \label{pNdot1y}
\end{equation}
\begin{equation}
\frac{\partial ^{3}}{\partial t^{2}~\partial z}\left[ z^{\bar{m}%
-N-1}~p_{N+1}\left( z;t\right) \right] =\sum_{m=1,~m\neq \bar{m}}^{N+1}\left[
\left( \bar{m}-m\right) ~\ddot{y}_{m}  ~z^{\bar{m}-m-1}\right]
~;  \label{pNdot2y}
\end{equation}%
note that here we again emphasized---by excluding from the sums in the
right-hand sides the (vanishing!) term with $m=\bar{m}$---the obvious fact
that these formulas are \textit{independent} of the function $y_{\bar{m}}\left(
t\right) $. And clearly these formulas, when evaluated at $z=x_{n}\left(
t\right) $, read---for all values of $n=1,2,...,N$---as follows: 
\end{subequations}
\begin{subequations}
\begin{eqnarray}
&&\left. \left\{ \frac{\partial ^{2}}{\partial t~\partial z}\left[ z^{\bar{m}%
-N-1}~p_{N+1}\left( z;t\right) \right] \right\} \right\vert _{z=x_{n}  }  \notag \\
&=&\sum_{m=1,~m\neq \bar{m}}^{N+1}\left[ \left( \bar{m}-m\right) ~\dot{y}%
_{m}  ~\left( x_{n}  \right) ^{\bar{m}-m-1}%
\right] ~,  \label{dtdzZPy}
\end{eqnarray}

\begin{eqnarray}
&&\left. \left\{ \frac{\partial ^{3}}{\partial t^{2}~\partial z}\left[ z^{%
\bar{m}-N-1}~p_{N+1}\left( z;t\right) \right] \right\} \right\vert
_{z=x_{n}  }  \notag \\
&=&\sum_{m=1,~m\neq \bar{m}}^{N+1}\left[ \left( \bar{m}-m\right) ~\ddot{y}%
_{m}  ~\left( x_{n}  \right) ^{\bar{m}-m-1}%
\right] ~.  \label{d2tdzZPy}
\end{eqnarray}

It is on the other hand easily seen that the expressions for the quantities
analogous to (\ref{dtdzZPy}) that instead follow from the expression (\ref%
{pN+1ztb}) of the polynomial $p_{N+1}\left( z;t\right) ,$ when evaluated at $%
z=x_{1} (t) $, read as follows: 
\end{subequations}
\begin{eqnarray}
&&\left. \left\{ \frac{\partial ^{2}}{\partial t~\partial z}\left[ z^{\bar{m}%
-N-1}~p_{N+1}\left( z;t\right) \right] \right\} \right\vert _{z=x_{1}  }  \notag \\
&=&-2~\left( x_{1}  \right) ^{\bar{m}-N-1}~\left[
\tprod\limits_{\ell =2}^{N}\left( x_{1}  -x_{\ell }  \right) \right] ~\dot{x}_{1}  ~.  \label{dtdzZPx1}
\end{eqnarray}

Equating these equations to (\ref{pNdot1y})---also evaluated at $%
z=x_{1}\left( t\right)  $---yields the sought expressions of the first
derivatives $\dot{x}_{1}  $ of the $N$ zero $x_{1}\left(
t\right) $: 
\begin{equation}
\dot{x}_{1}  =\left[ 2~\tprod\limits_{\ell =2}^{N}\left(
x_{1}  -x_{\ell }  \right) \right]
^{-1}~\sum_{m=1,~m\neq \bar{m}}^{N+1}\left[ \left( m-\bar{m}\right) ~\dot{y}%
_{m}  ~\left( x_{1}  \right) ^{N-m}\right] ~.
\end{equation}

In an analogous manner the equations are obtained for the second $t$%
-derivatives of the $N$ zeros $x_{n}\left( t\right) $ (a check of their
derivation is left to the willing reader). They read as follows: 
\begin{subequations}
\label{x1ndotdot}
\begin{eqnarray}
\ddot{x}_{1}&=&-\left( N+1-\bar{m}\right) ~\frac{\left( \dot{x}_{1}\right)
^{2}}{x_{1}}+\dot{x}_{1}~\sum_{n=2}^{N}~\left( \frac{2~\dot{x}_{n}+\dot{x}%
_{1}}{x_{1}-x_{n}}\right)   \nonumber \\
&&+\left[ 2~\tprod\limits_{n=2}^{N}\left( x_{1}-x_{n}\right) \right]
^{-1}\sum_{m=1,~m\neq \bar{m}}^{N+1}\left[ \left( m-\bar{m}\right) ~\ddot{y}%
_{m}~\left( x_{1}\right) ^{N-m}\right] ~, \label{x1dotdot}\\
\ddot{x}_{n}&=&\frac{2~\dot{x}_{1}~\dot{x}_{n}}{x_{n}-x_{1}}+\sum_{\ell
=1,~\ell \neq n}^{N}\left( \frac{2~\dot{x}_{n}~\dot{x}_{\ell }}{%
x_{n}-x_{\ell }}\right) \notag\\
&&+\frac{2~\left( \dot{x}_{1}\right) ^{2}}{x_{n}-x_{1}}%
~\left( \frac{x_{n}}{x_{1}}\right) ^{N+1-\bar{m}}\dprod\limits_{\ell
=2\,~\ell \neq n}^{N}\left( \frac{x_{1}-x_{\ell }}{x_{n}-x_{\ell }}\right) \notag\\
&&-\left[ \left( x_{n}\right) ^{\bar{m}-N-1}~\dprod\limits_{\ell =1,~\ell \neq
n}^{N}\left( x_{n}-x_{\ell }\right) \right] ^{-1} \notag \\
&& 
 \cdot \sum_{m=1,m\neq \bar{m}}^{N+1}\left\{ \ddot{y}_{m}~\left[ \frac{\left(
x_{n}\right) ^{\bar{m}-m}-\left( x_{1}\right) ^{\bar{m}-m}}{x_{n}-x_{1}}%
\right] \right\} ~,~~~n=2,...,N~.   \label{xndotdot}
\end{eqnarray}
In (\ref{xndotdot}) the quantity $\left[ \left( x_{n}\right) ^{\bar{m}%
-m}-\left( x_{1}\right) ^{\bar{m}-m}\right] /\left( x_{n}-x_{1}\right) $ can
of course be replaced using the identity 
\end{subequations}
\begin{subequations}
\begin{equation}
\frac{\left( x_{n}\right) ^{\bar{m}-m}-\left( x_{1}\right) ^{\bar{m}-m}}{%
x_{n}-x_{1}}=\sum_{j=0}^{\bar{m}-m-1}\left[ \left( x_{n}\right) ^{\bar{m}%
-m-1-j}~\left( x_{1}\right) ^{j}\right] ~~~\text{if}~~~m<\bar{m}~,
\end{equation}%
\begin{equation}
\frac{\left( x_{n}\right) ^{\bar{m}-m}-\left( x_{1}\right) ^{\bar{m}-m}}{%
x_{n}-x_{1}}=-\sum_{j=0}^{m-\bar{m}-1}\left[ \left( x_{n}\right) ^{-\left(
j+1\right) }~\left( x_{1}\right) ^{\bar{m}-m+j}\right] ~~~\text{if}~~~m>\bar{%
m}~.
\end{equation}%
And let us reemphasize that in these formulas, (\ref{x1ndotdot}), the
contribution of the coefficient $y_{m}$ with $m=\bar{m}$ is \textit{not}
present.

The idea is now to identify and investigate the $N+1$ dynamical systems
satisfied by the $N$ zeros $x_{n}\left( t\right) $ with $n=1,2,...,N$%
---explicitly yielded in an obvious manner by the equations written above,
together with the explicit equations expressing the $N+1$ coefficients $%
y_{m}\left( t\right) $ in terms of the $N$ zeros $x_{n}$  (see for instance (%
\ref{ymxSect22}))---which correspond
to ``solvable'' dynamical systems satisfied by the $N$ coefficients $%
y_{m}\left( t\right) $ with $m=1,2,...,\bar{m}-1,\bar{m}+1,...,N+1,$ for
every assigned value of the index $\bar{m}$ in the range from $1$ to $N+1$.
These systems satisfied by the $N$ zeros $x_{n}\left( t\right) $ with $%
n=1,2,...,N$ are then as well solvable by algebraic operations. Let us
indicate what the corresponding procedure is.

\textit{Step (i)}. Given the $N$ initial values $x_{n}\left( 0\right) $ and the $N$ initial
velocities $\dot{x}_n(0)$ of the
dynamical system satisfied by the $N$ zeros $x_{n}\left( t\right) ,$ compute
the $N+1$ initial values $y_{m}\left( 0\right) $ and the $N+1$ initial velocities $\dot{y}_m(0)$ via (\ref{ymxSect22}) (at $%
t=0$).

\textit{Step (ii)}. Compute $y_{m}\left( t\right) $ with $m=1,2,...,\bar{m}-1,\bar{m}%
+1,...,N+1$ by solving the---assumedly solvable---$N$ evolution equations
satisfied by these $N$ quantities (with
the initial values obtained from \textit{Step(1)}).

\textit{Step (iii)}. Note that, because the polynomial $p_{N+1}\left( z;t\right) $
features a double zero at $z=x_{1}\left( t\right) ,$ see (\ref{pN+1ztb}),
the function 
\end{subequations}
\begin{subequations}
\begin{equation}
\frac{\partial }{\partial z}\left[ z^{\bar{m}-N-1}~p_{N+1}\left( z;t\right) %
\right] =\sum_{m=1,~m\neq \bar{m}}^{N+1}\left[ \left( \bar{m}-m\right)
~y_{m}  ~z^{\bar{m}-m-1}\right]
\end{equation}%
vanishes at $z=x_{1}\left( t\right) ,$ hence 
\end{subequations}
\begin{subequations}
\begin{equation}
\sum_{m=1,~m\neq \bar{m}}^{N+1} \left[ \left( \bar{m}-m\right)
~y_{m}  ~\left( x_{1}  \right) ^{\bar{m}-m-1}%
\right]  =0~,
\end{equation}%
or equivalently (after multiplication by $\left[ x_{1}  \right]
^{N-\bar{m}+2}$), 
\end{subequations}
\begin{subequations}
\begin{equation}
\sum_{m=1,~m\neq \bar{m}}^{N+1} \left[ \left( \bar{m}-m\right)
~y_{m}  ~\left( x_{1}  \right) ^{N+1-m}\right]
 =0~.
\end{equation}%
Note that this is, \textit{de facto,} an algebraic equation of degree $N$
for the quantity $x_{1}\left( t\right) ,$ from which this quantity can be
computed for all time $t$ (since all the quantities $y_{m}\left( t\right) $
with $m\neq \bar{m}$ have been evaluated at \textit{Step (ii)}---and note that indeed
the quantity $y_{\bar{m}}\left( t\right) $ does not appear in this algebraic
equation). In this manner, by the algebraic operation of finding the roots
of a polynomial of degree $N+1$, one can in principle obtain the quantity $%
x_{1}\left( t\right) $ for all time. In fact, one obtains generally $N+1$
values of this quantity for all values of $t,$ but by following---by
continuity in $t$---these $N$ values all the way back to $t=0$ one can
identify the solution $x_{1}\left( t\right) $ as the one that yields at $t=0$
the assigned initial value $x_{1}\left( 0\right) .$ So \textit{Step (iii)} allows to
identify---by algebraic operations---the solution $x_{1}\left( t\right) $
for all time.

\textit{Step (iv)}. It is plain (see (\ref{pN+1zta})) that 
\end{subequations}
\begin{subequations}
\begin{equation}
\left( x_{1}  \right) ^{N+1}+\sum_{m=1}^{N+1}\left[
y_{m}  ~\left( x_{1}  \right) ^{N+1-m}\right]
=0~,
\end{equation}%
hence%
\begin{equation}
y_{\bar{m}}  =-\left( x_{1}  \right) ^{\bar{m}%
}-\sum_{m=1,~m\neq m}^{N+1}\left[ y_{m}  ~\left( x_{1}  \right) ^{\bar{m}-m}\right] ~.
\end{equation}%
\end{subequations}
This shows that $y_{\bar{m}}  $ is now also known for all time.

\textit{Step (v)}. Finally, from the knowledge of \textit{all} the $N+1$ coefficients 
$y_{m}  ,$ the $N$ zeros of the polynomial (\ref{pN+1zta}) can
be obtained via an algebraic operation, completing the task to solve the
dynamical system characterizing the time evolution of the $N$ zeros $%
x_{n}  $.

In an actual numerical implementation of this procedure the accuracy with
which one of the zeros of this polynomial would turn out to be \textit{double%
} (and therefore identified as $x_{1}\left( t\right) $), and the discrepancy
of the values of this double zero from the value of $x_{1}\left( t\right) $
computed in \textit{Step (iii)}, would provide an estimate of the numerical precision
of the treatment. Moreover---and perhaps more importantly---the fact should
be re-emphasized that the $N+1$ \textit{different} dynamical systems satisfied by the $N$ zeros 
$x_{n}\left( t\right) $---corresponding to the $N+1$ assignments of the
index $\bar{m}$ in the range from $1$ to $N+1$, see above---shall \textit{%
all inherit} the properties of the system of $N$ evolution equations satisfied
by the $N$ coefficients $y_{m}\left( t\right) $ with $m=1,2,...,\bar{m}-1,%
\bar{m}+1,...N+1$ (see examples below).

\bigskip

\section{New systems of solvable nonlinear evolution equations}

\bigskip

In this Section we illustrate the findings of Section~2 by several examples of \textit{new solvable} $2$ or $3$-body 
problems obtained from several simple yet representative models. 

\subsection{Example 3.1}

In this example, we take as a point of  departure one of the following $3$ \textit{generating}
models: 
\begin{equation}
\text{\textbf{Model 3.1.1}: }\ddot{y}_{1}=\mathbf{i~}r_{1}~\omega ~\dot{y}_{1},~~~\ddot{y}%
_{2}=\mathbf{i~}r_{2}~\omega ~\dot{y}_{2}~;  \label{Model3}
\end{equation}%
\begin{equation}
\text{\textbf{Model 3.1.2}: } \ddot{y}%
_{1}=\mathbf{i~}r_{1}~\omega ~\dot{y}_{1},~~~  \ddot{y}_{3}=\mathbf{i~}r_{3}~\omega ~\dot{y}_{3}~;  \label{Model2}
\end{equation}%
\begin{equation}
\text{\textbf{Model 3.1.3}: }\ddot{y}_{2}=\mathbf{i~}r_{2}~\omega ~\dot{y}_{2},~~~\ddot{y}%
_{3}=\mathbf{i~}r_{3}~\omega ~\dot{y}_{3}~. \label{Model1}
\end{equation}
Here and hereafter $\omega $ is an arbitrary nonvanishing real
number; $r_{1},~r_{2},r_{3}$ are $3$ arbitrary nonvanishing rational
numbers; $\mathbf{i}$ is the imaginary unit, so that $\mathbf{i}%
^{2}=-1$. These 3 models are \textit{Hamiltonian} and \textit{integrable} and
their solutions
\begin{equation}
y_{m}\left( t\right) =y_{m}\left( 0\right) +\dot{y}_{m}\left( 0\right) ~%
\left[ \frac{\exp \left( \mathbf{i}~r_{m}~\omega ~t\right) -1}{\mathbf{i}%
~r_{m}~\omega }\right] ~,~~~m=1,2,3~,
\label{ymModels1}
\end{equation}
are \textit{isochronous} with a period which is an \textit{integer multiple}
of the basic period%
\begin{equation}
T=\frac{2~\pi }{\left\vert \omega \right\vert }~.  \label{T}
\end{equation}

\textbf{Remark 3.1.1.} The fact that the last three models are Hamiltonian follows from the observation that for every complex $\alpha$, the equation $\ddot{y}=\alpha ~\dot{y}$ is generated by the Hamiltonian $H(y,p)=e^p- \alpha~ y$. $\blacksquare$

These generating models yield the following solvable two-body problems, via the method described in Subsection 2.1, see~(\ref{xdotdot1}),~(\ref{xdotdot2})  and~(\ref{xdotdot3}):

\textbf{System 3.1.1:}
\begin{eqnarray}
\ddot{x}_1&=&\frac{1}{x_1-x_2}\Big\{
\dot{x}_1 \,\left[
-\mathbf{i} \,r_2\, \omega\, x_2+\dot{x}_1+2 \dot{x}_2
 \right]\notag\\
 &&+\mathbf{i}\,\omega\, x_1\,\left[
 (2 \,r_1-r_2) \,\dot{x}_1 +(r_1-r_2) \,\dot{x}_2
 \right]
\Big\},
 \notag\\
\ddot{x}_2&=&-\frac{\mathbf{i}}{x_1-x_2}\Big\{
-2\,\mathbf{i}\, \dot{x}_1\,(\dot{x}_1+2\dot{x}_2)+\omega\, x_2\,\left[
2\,(r_1-r_2)\, \dot{x}_1+r_1 \,\dot{x}_2
\right]
\notag\\
&&+\omega \,x_1 \left[
2\,(r_1-r_2)\,\dot{x}_1+(r_1-2\,r_2)\,\dot{x}_2
\right]
\Big\}.
\label{syst:Order2MGoldA}
\end{eqnarray}

\textbf{System 3.1.2:}
\begin{eqnarray}
\ddot{x}_1&=&\frac{1}{2 \,x_1 \,(x_1-x_2)}\Big\{
2\, x_2\, \dot{x}_1^2+2\, x_1\, \dot{x}_1\,\left[
-\mathbf{i}\, r_3\, \omega\, x_2+2\, \dot{x}_2
\right]\notag\\
&&+\mathbf{i}\,\omega\, (x_1)^2\, \left[
2 \,r_1\, \dot{x}_1+(r_1-r_3)\, \dot{x}_2
\right]
\Big\},
 \notag\\
\ddot{x}_2&=&\frac{\mathbf{i}}{x_1\,(x_1-x_2)}\Big\{
2\, \mathbf{i}\, x_2 \,(\dot{x}_1)^2+r_3\, \omega \,x_1^2 \,\dot{x}_2 +
4 \,\mathbf{i}\,x_1\, \dot{x}_1\, \dot{x}_2
\notag\\
&&-\omega\, x_1\, x_2\,\left[
2\,(r_1-r_3)\,\dot{x}_1+r_1\, \dot{x}_2
\right]
\Big\}.
\label{syst:Order2MGoldB}
\end{eqnarray}

\textbf{System 3.1.3:}
\begin{eqnarray}
\ddot{x}_1&=&\frac{\mathbf{i}}{x_1\,(x_1-x_2)}\Big\{
-2\,\mathbf{i}\, x_2 \,(\dot{x}_1)^2+x_1\, \dot{x}_1
\left[
(r_2-2\,r_3)\,\omega\, x_2+\mathbf{i}\, (\dot{x}_1-2\,\dot{x}_2)
\right]\notag\\
&&
+\omega\, (x_1)^2\left[
r_2 \,\dot{x}_1+(r_2-r_3)\,\dot{x}_2
\right]
\Big\},
\notag\\
\ddot{x}_2&=&\frac{\mathbf{i}}{(x_1)^2 (x_1-x_2)} \Big\{
2\,(-r_2+r_3)\, \omega \,x_1\, (x_2)^2 \,\dot{x}_1 +2\,\mathbf{i}\, (x_2)^2 \,(\dot{x}_1)^2
+r_3\,\omega \,(x_1)^3 \,\dot{x}_2\notag\\
&&+
4\,\mathbf{i}\, (x_1)^2 \,\dot{x}_1 \,\dot{x}_2
+\omega \,(x_1)^2 \, x_2\,\left[
-2\,(r_2-r_3)\,\dot{x}_1+(-2\, r_2+r_3)\, \dot{x}_2
\right]
\Big\}.
\label{syst:Order2MGoldC}
\end{eqnarray}
These 3 systems are Hamiltonian,  solvable by algebraic operations---which in these cases might even be
performed explicitly, although the resulting formulas, including quadratic and
cubic roots, would hardly be enlightening---and their solutions are 
\textit{isochronous}. 

Below we provide the plots of the solutions of system~(\ref{syst:Order2MGoldA}) with the parameters
\begin{equation}
r_1=\frac{1}{2},\; \; r_2=\frac{1}{3}, \;\;\omega=2\pi,
\label{par:Order2MGoldA}
\end{equation}
satisfying the initial conditions
\begin{eqnarray}
&&x_1(0) =0.90 - 0.19 \,\mathbf{i}, \;\;x_1'(0) = 0.085 - 0.37 \,\mathbf{i}, \notag\\
&&x_2(0) =1.96 + 1.75 \,\mathbf{i}, \;\;x_2'(0) = -0.34 + 2.14 \,\mathbf{i}.
\label{InitCond:Order2MGoldA}
\end{eqnarray}

\begin{minipage}{\linewidth}
      \centering
      \begin{minipage}{0.45\linewidth}
          \begin{figure}[H]
              \includegraphics[width=\linewidth]{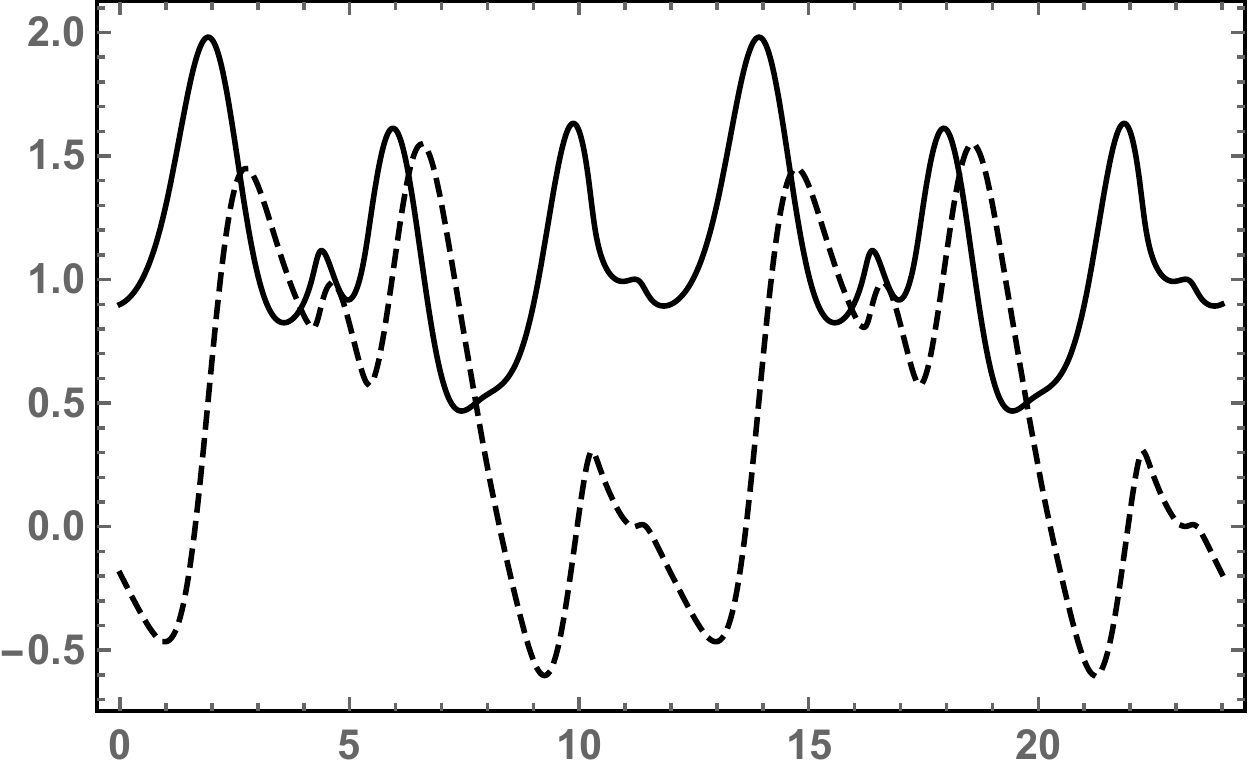}
              \caption{Initial value problem~(\ref{syst:Order2MGoldA}),~(\ref{par:Order2MGoldA}),~(\ref{InitCond:Order2MGoldA}). Graphs of the real (bold curve) and imaginary 
              (dashed curve) parts of the coordinate $x_1(t)$; period $12$.}
              \label{F1A1}
          \end{figure}
      \end{minipage}
      \hspace{0.05\linewidth}
      \begin{minipage}{0.45\linewidth}
          \begin{figure}[H]
              \includegraphics[width=\linewidth]{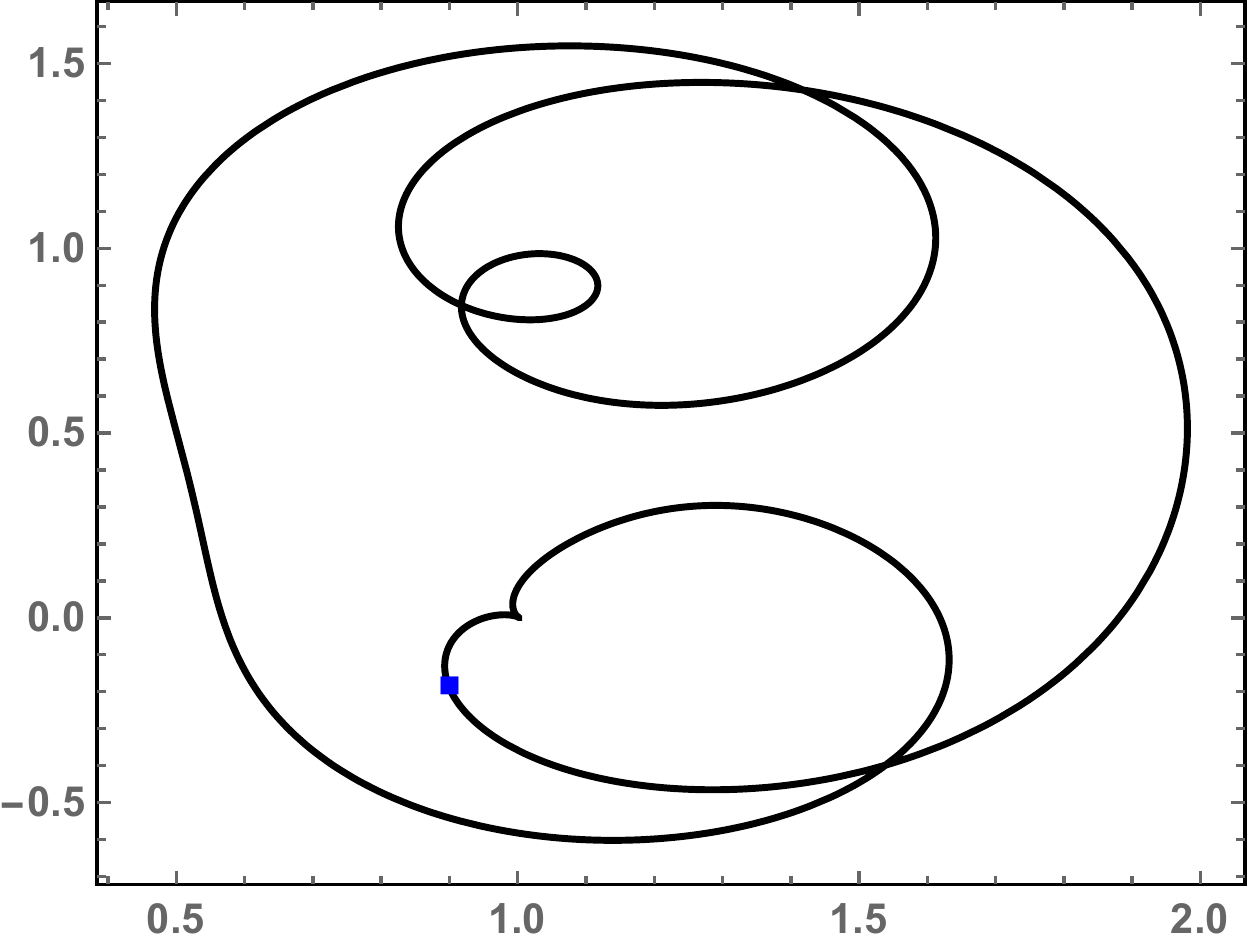}
              \caption{Initial value problem~(\ref{syst:Order2MGoldA}),~(\ref{par:Order2MGoldA}),~(\ref{InitCond:Order2MGoldA}). Trajectory, in the complex $x$-plane, of  $x_1(t)$; 
              period $12$. The   square indicates the initial condition $x_1(0)=0.90 - 0.19 \,\mathbf{i}$.}
              \label{F1A2}
          \end{figure}
      \end{minipage}
  \end{minipage}
  
  \begin{minipage}{\linewidth}
      \centering
      \begin{minipage}{0.45\linewidth}
          \begin{figure}[H]
              \includegraphics[width=\linewidth]{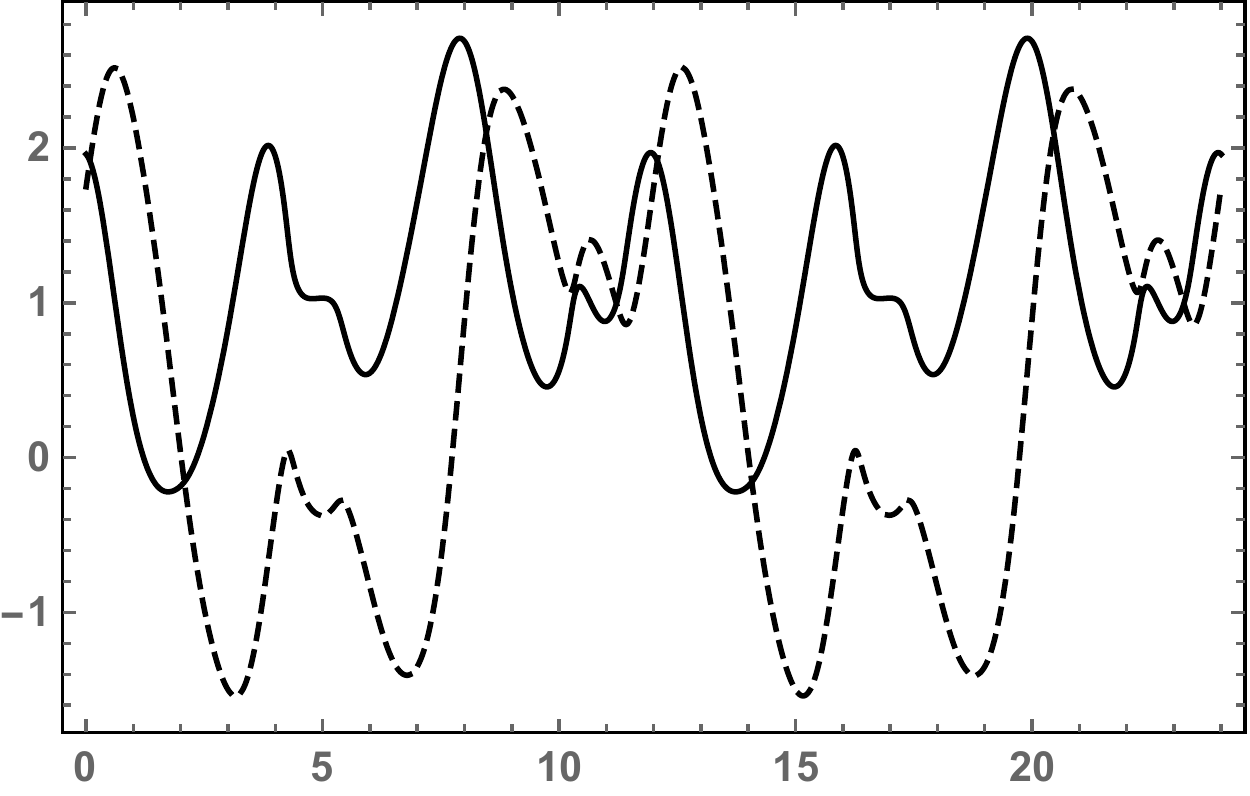}
              \caption{Initial value problem~(\ref{syst:Order2MGoldA}),~(\ref{par:Order2MGoldA}),~(\ref{InitCond:Order2MGoldA}). Graphs of the real (bold curve) and imaginary 
              (dashed curve) parts of the coordinate $x_2(t)$; period $12$.}
              \label{F1A3}
          \end{figure}
      \end{minipage}
      \hspace{0.05\linewidth}
      \begin{minipage}{0.45\linewidth}
          \begin{figure}[H]
              \includegraphics[width=\linewidth]{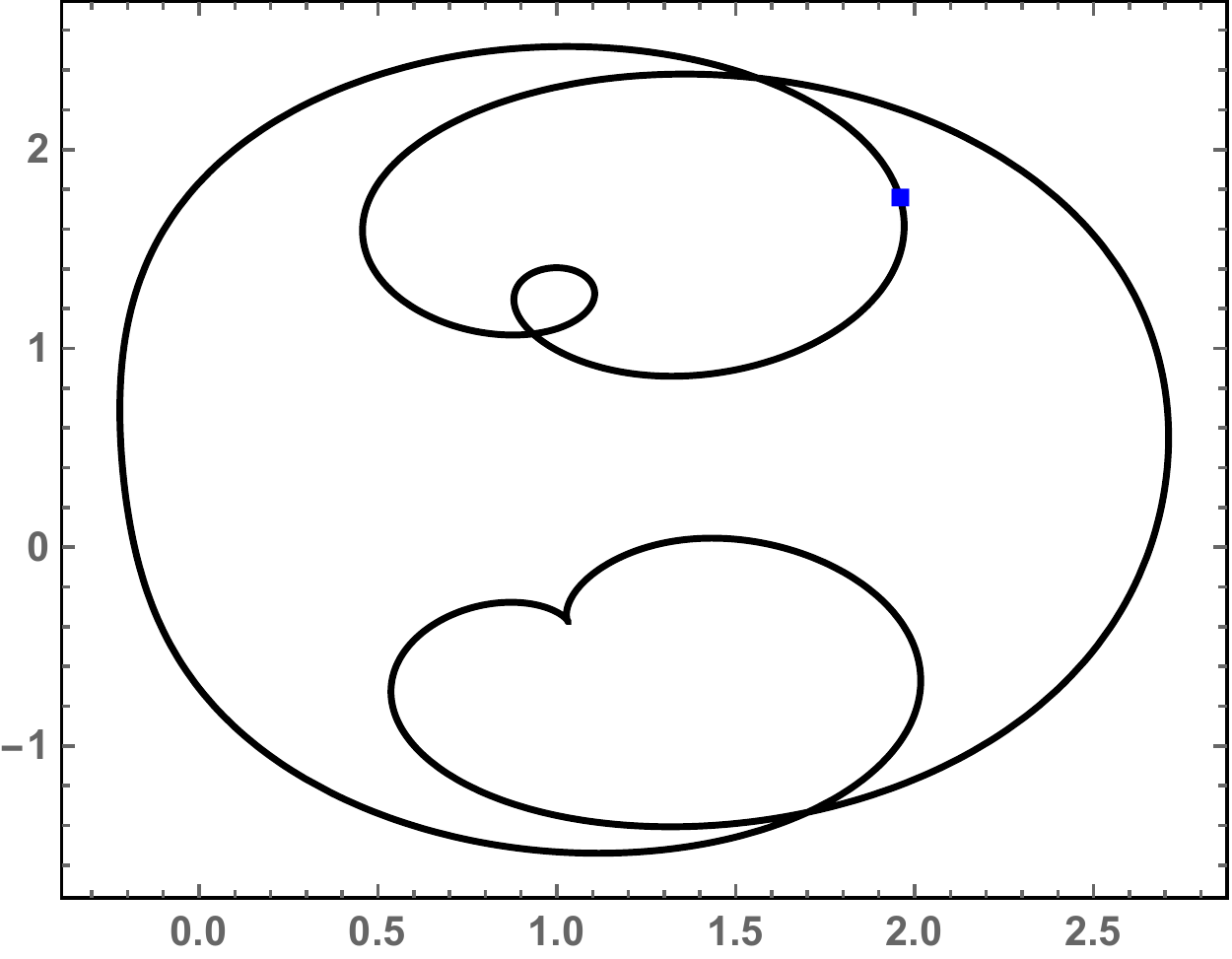}
              \caption{Initial value problem~(\ref{syst:Order2MGoldA}),~(\ref{par:Order2MGoldA}),~(\ref{InitCond:Order2MGoldA}). Trajectory, in the complex $x$-plane, of  $x_2(t)$; 
              period $12$. The   square indicates the initial condition $x_2(0)= 1.96 + 1.75 \,\mathbf{i}$.}
              \label{F1A4}
          \end{figure}
      \end{minipage}
  \end{minipage}
\smallskip

\textbf{Remark 3.1.2}. The reader who wonders why the period of the solution of the initial value problem~(\ref{syst:Order2MGoldA}),~(\ref{par:Order2MGoldA}),~(\ref{InitCond:Order2MGoldA}) is $12$ rather than $6$ is advised to read Ref.~\cite{18}.
$\blacksquare$

\textbf{Remark 3.1.3}. Equations of motion~(\ref{syst:Order2MGoldA}), (\ref{syst:Order2MGoldB}), (\ref{syst:Order2MGoldC}) drastically simplify in the
special case with $r_{1}=r_{2}=r_3=r,$ when they read

\begin{eqnarray}
&&\ddot{x}_{1}=\left( x_{1}-x_{2}\right) ^{-1}~\dot{x}_{1}~\left( \dot{x}%
_{1}+2~\dot{x}_{2}\right) +\mathbf{i}~\omega ~r~\dot{x}_{1}~,\notag\\
&&\ddot{x}_{2}=-2~\left( x_{1}-x_{2}\right) ^{-1}~\dot{x}_{1}~\left( \dot{x}%
_{1}+2~\dot{x}_{2}\right) +\mathbf{i}~\omega ~r~\dot{x}_{2}~;
\end{eqnarray}%
\begin{eqnarray}
&&\ddot{x}_{1}=\left[ x_{1}~\left( x_{1}-x_{2}\right) \right] ^{-1}~\dot{x}%
_{1}~\left( \dot{x}_{1}~x_{2}+2~\dot{x}_{2}~x_{1}\right) +\mathbf{i}~\omega
~r~\dot{x}_{1}~,\notag\\
&&\ddot{x}_{2}=-2~\left[ x_{1}~\left( x_{1}-x_{2}\right) \right] ^{-1}~\dot{x}%
_{1}~\left( \dot{x}_{1}~x_{2}+2~\dot{x}_{2}~x_{1}\right) +\mathbf{i}~\omega
~r~\dot{x}_{2}~;
\end{eqnarray}%
\begin{eqnarray}
&&\ddot{x}_{1}=-\left[ x_{1}~\left( x_{1}-x_{2}\right) \right] ^{-1}~\dot{x}%
_{1}~\left[ \dot{x}_{1}~\left( x_{1}-2~x_{2}\right) -2~\dot{x}_{2}~x_{1}%
\right] +\mathbf{i}~\omega ~r~\dot{x}_{1}~,\notag\\
&&\ddot{x}_{2}=-2~\left[ x_{1}^{2}~\left( x_{1}-x_{2}\right) %
\right] ^{-1}~\dot{x}_{1}~\left[ \dot{x}_{1}~x_{2}^{2}+2~\dot{%
x}_{2}~ x_{1}^{2}\right] +\mathbf{i}~\omega ~r~\dot{x}%
_{2}~.
\end{eqnarray}
\vspace{-3mm}
$\blacksquare$

\subsection{Example 3.2}

In this example, we consider solvable $2$-body problems generated by the following $3$ models: 
\begin{equation}
\text{\textbf{Model 3.2.1}: }\ddot{y}_{1}=-r_{1}^2~\omega^2 ~{y}_{1},~~~\ddot{y}%
_{2}=-r_{2}^2~\omega^2 ~{y}_{2}~;  \label{Model21}
\end{equation}%
\begin{equation}
\text{\textbf{Model 3.2.2}: } \ddot{y}%
_{1}=-r_{1}^2~\omega^2~{y}_{1},~~~  \ddot{y}_{3}=-r_{3}^2~\omega^2~{y}_{3}~;  \label{Model22}
\end{equation}%
\begin{equation}
\text{\textbf{Model 3.2.3}: }\ddot{y}_{2}=-r_{2}^2~\omega^2 ~{y}_{2},~~~\ddot{y}%
_{3}=-r_{3}^2~\omega^2~{y}_{3}~. \label{Model23}
\end{equation}
Similarly to Example~1, $\omega $ is an arbitrary nonvanishing real
number; and $r_{1},~r_{2},r_{3}$ are $3$ arbitrary nonvanishing rational
numbers. These 3 models are \textit{Hamiltonian} and \textit{integrable}
and their solutions
\begin{equation}
y_{m}\left( t\right) =y_{m}\left( 0\right) \cos (r_m~\omega~t)+\frac{1}{r \omega}\dot{y}_{m}\left( 0\right) ~\sin(r_m~\omega~t)~,~~~m=1,2,3~,
\label{ymModels2}
\end{equation}
are \textit{isochronous} with a period which is an \textit{integer multiple}
of the basic period~(\ref{T}).

\textbf{Remark 3.2.1}. The last three systems are Hamiltonian because for every complex $\alpha$,  the equation $\ddot{y}=\alpha ~{y}$ is produced by the Hamiltonian $H(y,p)=p^2/2-\alpha~ y^2/2$. $\blacksquare$

The following  two-body problems are generated by the method described in Subsection 2.1, see~(\ref{xdotdot1}),~(\ref{xdotdot2})  and~(\ref{xdotdot3}):

\textbf{System 3.2.1:}
\begin{eqnarray}
&&\ddot{x}_1=\frac{
 \omega^2 x_1\left[(-4 r_1^2+r_2^2) x_1+2 (-r_1^2+r_2^2) x_2 \right]+2 \dot{x}_1(\dot{x}_1+2 \dot{x}_2)
}{2 (x_1-x_2)}, \notag\\
&&\ddot{x}_2=\frac{
\omega^2 x_1\left[ (2 r_1^2-r_2^2) x_1+(3 r_1^2-2 r_2^2) x_2 \right] + r_1^2 \omega^2 x_2^2 +2 \dot{x}_1(\dot{x}_1+2 \dot{x}_2)
}
{(x_1-x_2)}.
\label{syst:Order2HarmOscA}
\end{eqnarray}

\textbf{System 3.2.2:}
\begin{eqnarray}
&&\ddot{x}_1=\frac{
 \omega^2 x_1^2\left[ -2 r_1^2 x_1+(-r_1^2+r_3^2) x_2 \right]+2\dot{x}_1( x_2 \dot{x}_1 +2 x_1 \dot{x}_2  )
}{2 x_1 (x_1-x_2)}, \notag\\
&&\ddot{x}_2=\frac{
\omega^2 x_1 x_2\left[ r_3^2 x_1-r_1^2(2 x_1+x_2) \right]+2 \dot{x}_1(x_2 \dot{x}_1+2 x_1 \dot{x}_2)
}
{x_1(x_1-x_2)}.
\label{syst:Order2HarmOscB}
\end{eqnarray}

\textbf{System 3.2.3:}
\begin{eqnarray}
&&\ddot{x}_1=\frac{
 \omega^2 x_1^2\left[ -r_2^2 x_1+2(-r_2^2+r_3^2)x_2 \right] +  2 \dot{x}_1\left[ 2 x_2 \dot{x}_1- x_1 (\dot{x}_1-2\dot{x}_2)\right ]
}{2 x_1 (x_1-x_2)}, \notag\\
&&\ddot{x}_2=\frac{
\omega^2 x_1^2 x_2 \left[ (r_2^2-r_3^2) x_1+(2 r_2^2-r_3^2) x_2 \right]-2\dot{x}_1\left[ x_2^2 \dot{x}_1 +2 x_1^2 \dot{x}_2 \right]
}
{x_1^2(x_1-x_2)}.
\label{syst:Order2HarmOscC}
\end{eqnarray}
These 3 systems are Hamiltonian,  solvable by algebraic operations and their solutions are 
\textit{isochronous}. 

Below we provide the plots of the solutions of system~(\ref{syst:Order2HarmOscA}) with the parameters
\begin{equation}
r_1=\frac{1}{2},\; \; r_2=\frac{1}{3}, \;\;\omega=2\pi,
\label{par:Order2HarmOscA}
\end{equation}
satisfying the initial conditions
\begin{eqnarray}
&&x_1(0) = -2.4 - 1.21 \,\mathbf{i}, \;\;x_1'(0) = -6.82 - 3.92 \,\mathbf{i}, \notag\\
&&x_2(0) = 4.89 + 2.42  \,\mathbf{i}, \;\;x_2'(0) = -6.81 - 2.44 \,\mathbf{i}.
\label{InitCond:Order2HarmOscA}
\end{eqnarray}

\begin{minipage}{\linewidth}
      \centering
      \begin{minipage}{0.45\linewidth}
          \begin{figure}[H]
              \includegraphics[width=\linewidth]{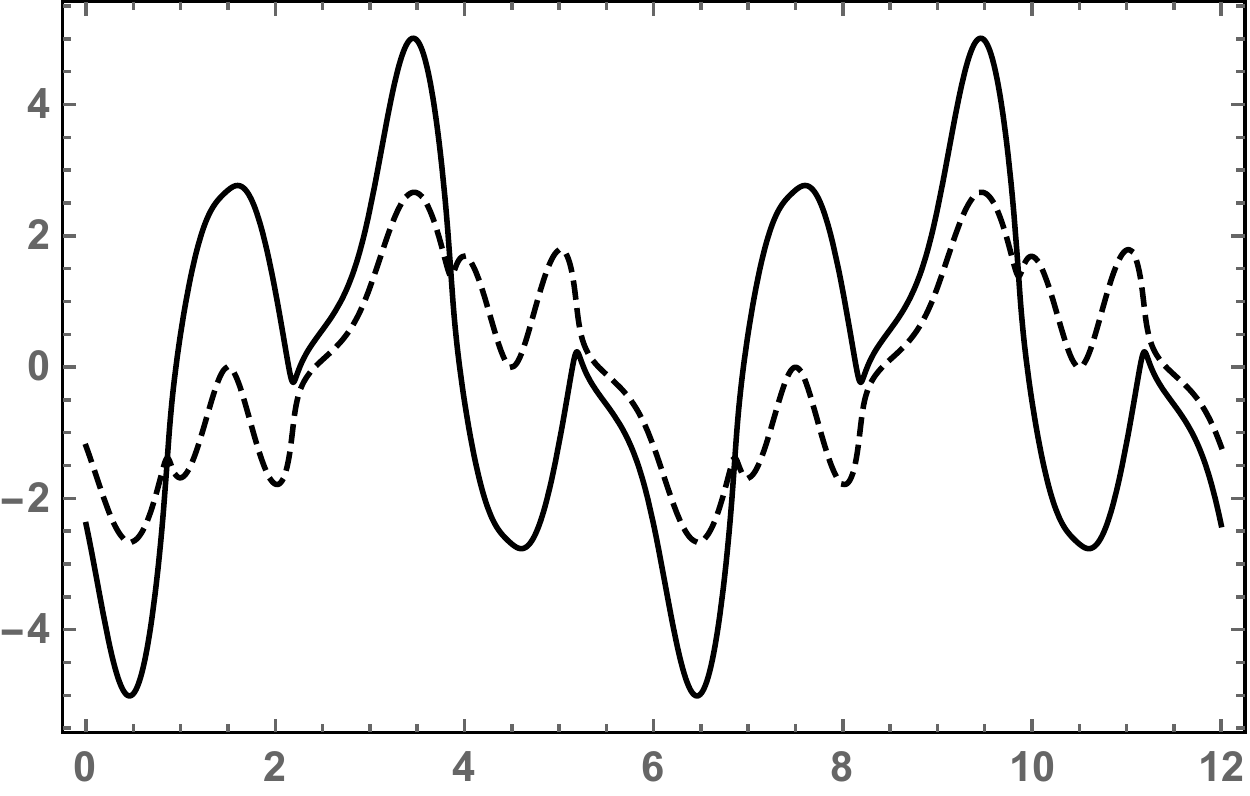}
              \caption{Initial value problem~(\ref{syst:Order2HarmOscA}),~(\ref{par:Order2HarmOscA}),~(\ref{InitCond:Order2HarmOscA}). Graphs of the real (bold curve) and imaginary 
              (dashed curve) parts of the coordinate $x_1(t)$; period $6$.}
              \label{F1A1}
          \end{figure}
      \end{minipage}
      \hspace{0.05\linewidth}
      \begin{minipage}{0.45\linewidth}
          \begin{figure}[H]
              \includegraphics[width=\linewidth]{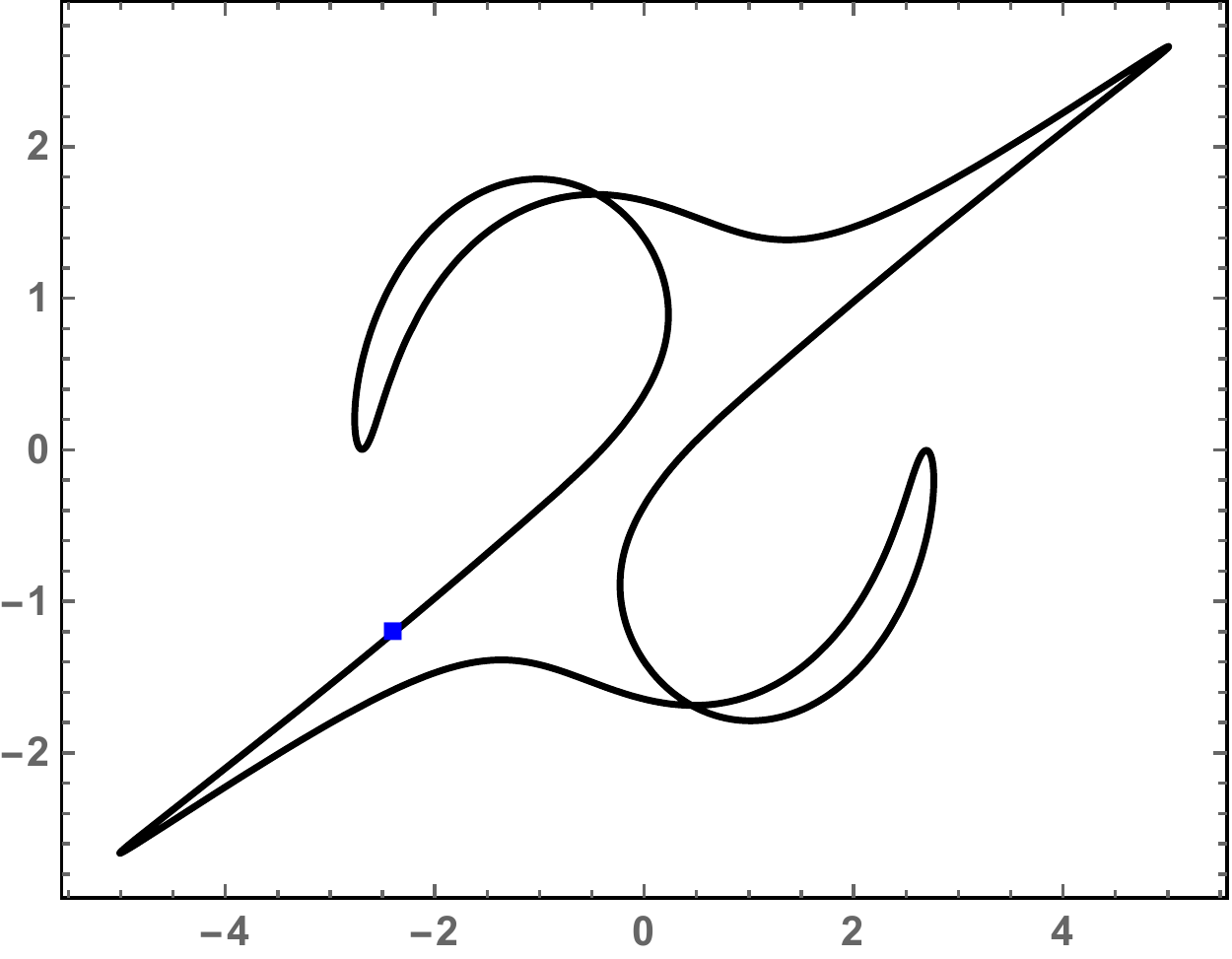}
              \caption{Initial value problem~(\ref{syst:Order2HarmOscA}),~(\ref{par:Order2HarmOscA}),~(\ref{InitCond:Order2HarmOscA}). Trajectory, in the complex $x$-plane, of  $x_1(t)$; 
              period $6$. The   square indicates the initial condition $x_1(0)=-2.4 - 1.21 \,\mathbf{i}$.}
              \label{F1A2}
          \end{figure}
      \end{minipage}
  \end{minipage}
  
  \begin{minipage}{\linewidth}
      \centering
      \begin{minipage}{0.45\linewidth}
          \begin{figure}[H]
              \includegraphics[width=\linewidth]{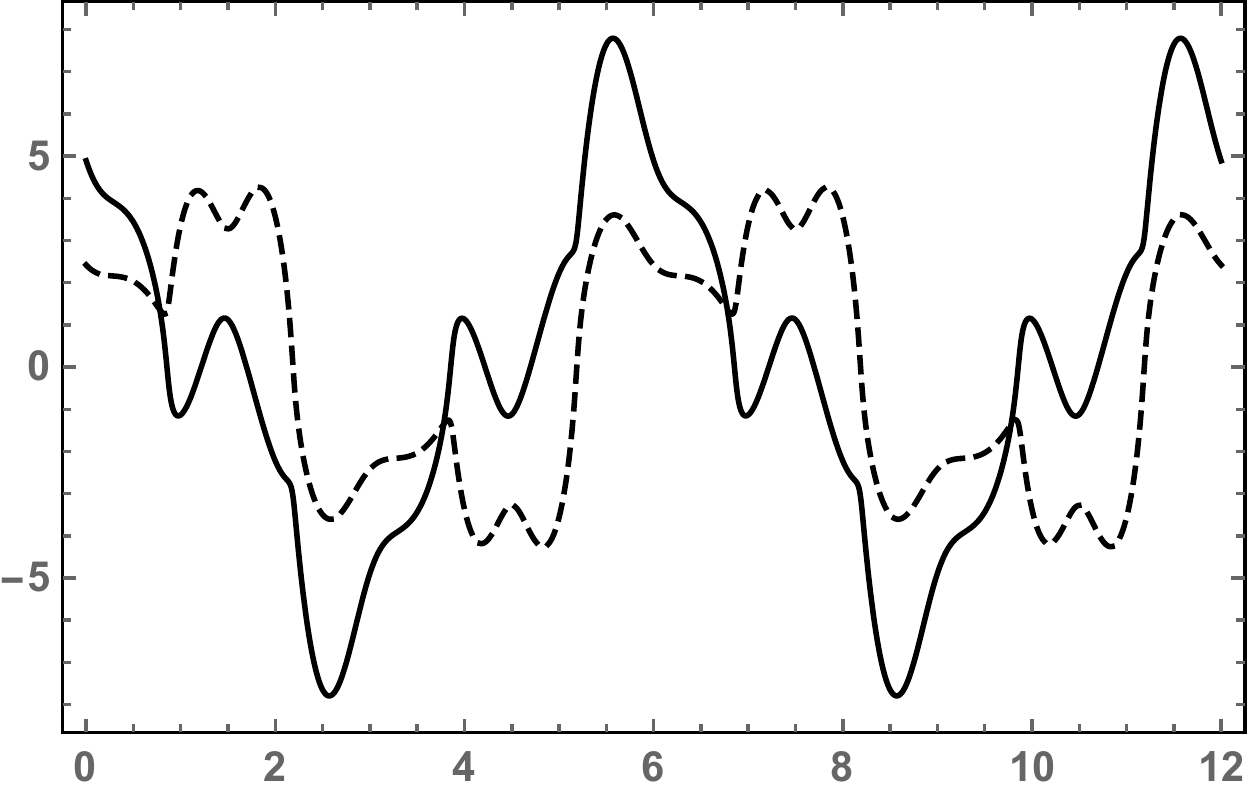}
              \caption{Initial value problem~(\ref{syst:Order2HarmOscA}),~(\ref{par:Order2HarmOscA}),~(\ref{InitCond:Order2HarmOscA}). Graphs of the real (bold curve) and imaginary 
              (dashed curve) parts of the coordinate $x_2(t)$; period $6$.}
              \label{F1A3}
          \end{figure}
      \end{minipage}
      \hspace{0.05\linewidth}
      \begin{minipage}{0.45\linewidth}
          \begin{figure}[H]
              \includegraphics[width=\linewidth]{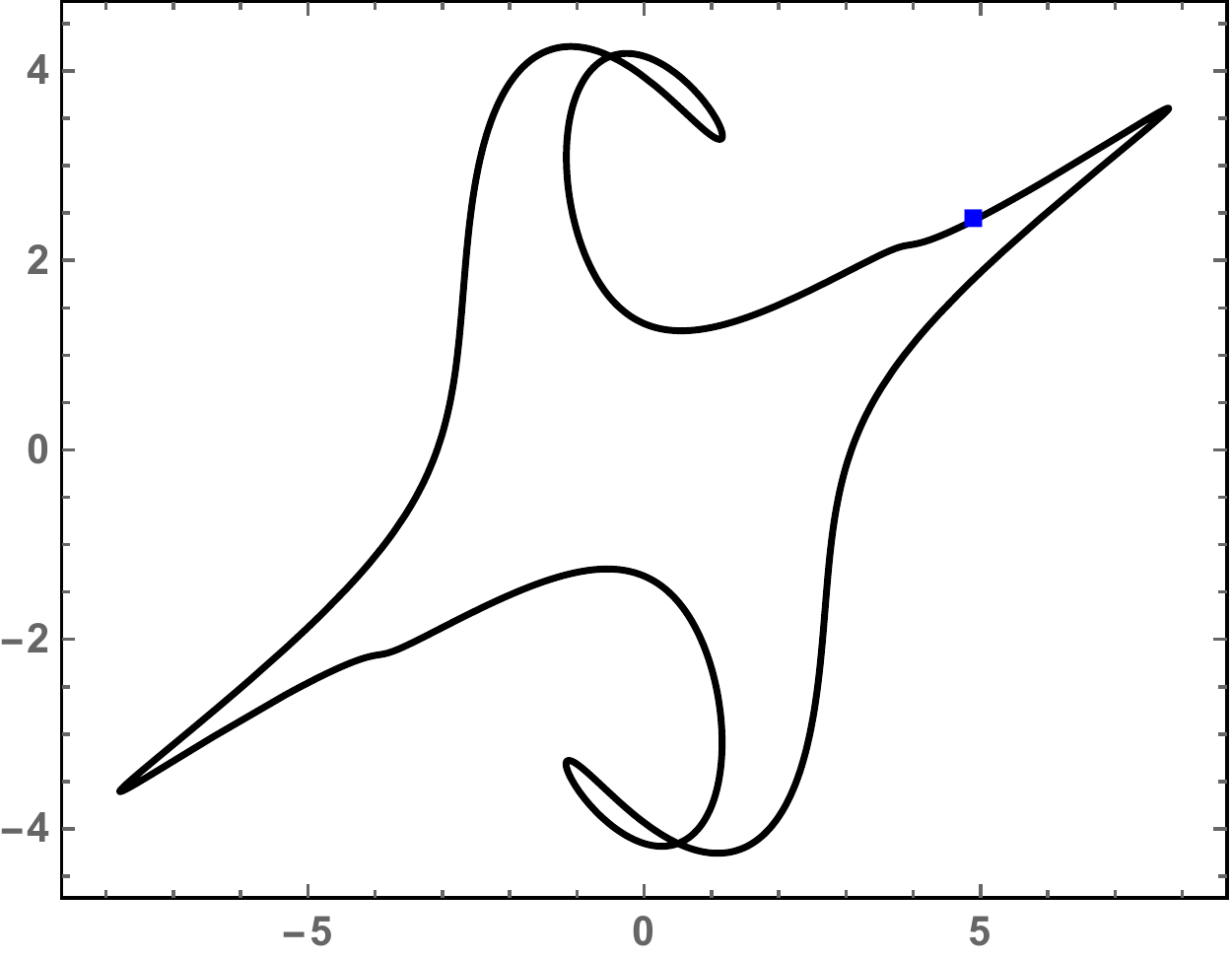}
              \caption{Initial value problem~(\ref{syst:Order2HarmOscA}),~(\ref{par:Order2HarmOscA}),~(\ref{InitCond:Order2HarmOscA}). Trajectory, in the complex $x$-plane, of  $x_2(t)$; 
              period $6$. The   square indicates the initial condition $x_2(0)= 4.89 + 2.42  \,\mathbf{i}$.}
              \label{F1A4}
          \end{figure}
      \end{minipage}
  \end{minipage}

\textbf{Remark 3.2.2}. Equations of motion~(\ref{syst:Order2HarmOscA}), (\ref{syst:Order2HarmOscB}), (\ref{syst:Order2HarmOscC}) simplify significantly  in the
special case where $r_{1}=r_{2}=r_3=r$:
\begin{eqnarray}
&&\ddot{x}_1=\left[ 2(x_1-x_2) \right]^{-1}
\left[ 
-3\omega^2 r^2 x_1^2+2\dot{x}_1(\dot{x}_1+2\dot{x}_2)
\right],\notag\\
&&\ddot{x}_2=(x_1-x_2)^{-1} \left[
\omega^2 r^2 (x_1^2+x_1 x_2+ x_2^2)+2 \dot{x}_1 (\dot{x}_1+2\dot{x}_2)
\right];
\end{eqnarray}
\begin{eqnarray}
&&\ddot{x}_1=\left[ 2 x_1 (x_1-x_2) \right]^{-1} 
\left[
-2\omega^2 r^2 x_1^3+2\dot{x}_1 (x_2 \dot{x}_1 +2 x_1 \dot{x}_2)
\right],\notag\\
&&\ddot{x}_2=\left[x_1(x_1-x_2) \right]^{-1} \left[
\omega^2 r^2 x_1 x_2 (x_2-x_1) +2\dot{x}_1(x_2 \dot{x}_1 +2 x_1 \dot{x}_2)
\right];
\end{eqnarray}
\begin{eqnarray}
&&\ddot{x}_1=\left[ 2 x_1 (x_1-x_2)\right]^{-1} \left[
-\omega^2 r^2 x_1^3 +2\dot{x}_1 (2 x_2 \dot{x}_1-x_1 \dot{x}_1+2 x_1\dot{x}_2)
\right],\notag\\
&&\ddot{x}_2=\left[ x_1^2 (x_1-x_2) \right]^{-1} \left[
\omega^2 r^2 x_1^2 x_2^2-2\dot{x}_1(x_2^2\dot{x}_1+2 x_1^2 \dot{x}_2)
\right].
\end{eqnarray}
\vspace{-3mm}
$\blacksquare$

\subsection{Example 3.3}

In this example, we generate \textit{solvable} $3$-body systems from the following $3$ 
models: 
\begin{equation}
\text{\textbf{Model 3.3.1}: }\ddot{y}_{1}=-r_{1}^2~\omega^2 ~{y}_{1},~~~\ddot{y}%
_{2}=\mathbf{i}~r_{2}~\omega~\dot{y}_{2}~;  \label{Model31}
\end{equation}%
\begin{equation}
\text{\textbf{Model 3.3.2}: } \ddot{y}%
_{1}=-r_{1}^2~\omega^2~{y}_{1},~~~  \ddot{y}_{3}=\mathbf{i}~r_{3}~\omega~\dot{y}_{3}~;  \label{Model32}
\end{equation}%
\begin{equation}
\text{\textbf{Model 3.3.3}: }\ddot{y}_{2}=-r_{2}^2~\omega^2~{y}_{2},~~~\ddot{y}%
_{3}=\mathbf{i}~r_{3}~\omega~\dot{y}_{3}~. \label{Model33}
\end{equation}
As in the previous Examples~3.1 and~3.2, $\omega $ is an arbitrary nonvanishing real
number and $r_{1},~r_{2},r_{3}$ are $3$ arbitrary nonvanishing rational
numbers. These 3 models are \textit{Hamiltonian} (see \textbf{Remarks 3.1.1} and \textbf{3.2.1}) and \textit{integrable}
and their solutions are given by appropriate combinations of 2 formulas chosen from among the 6 formulas~(\ref{ymModels1}) and~(\ref{ymModels2}). For example, the solution of Model~3.3.1 is given by~(\ref{ymModels2}) with $m=1$ and~(\ref{ymModels1}) with $m=2$. These models
are \textit{all isochronous} with a period which is an \textit{integer multiple}
of the basic period~(\ref{T}).

These generating models yield the following solvable two-body problems, via the method described in Subsection 2.1, see~(\ref{xdotdot1}),~(\ref{xdotdot2})  and~(\ref{xdotdot3}):

\textbf{System 3.3.1:}
\begin{eqnarray}
\ddot{x}_1&=&\frac{1}{x_1-x_2}\Big\{
\dot{x}_1 (\dot{x}_1+2\,\dot{x}_2)
-(r_1)^2 \,\omega^2 \,x_1\,(2\,x_1+x_2)
\notag\\
&&- \mathbf{i}\, r_2 \,\omega\, \left[ 
\dot{x}_1\,(x_1+x_2)+\dot{x}_2
\,x_1\right]
\Big\},
 \notag\\
\ddot{x}_2&=&\frac{1}{x_1-x_2} \Big\{
-2\,\dot{x}_1\,(\dot{x}_1+2\,\dot{x}_2)+(r_1)^2 \,\omega^2\,(x_1+x_2)\,(2\,x_1+x_2)\notag\\
&&+2 \,\mathbf{i}\, r_2\, \omega \left[
\dot{x}_1\,(x_1+x_2)+\dot{x}_2\, x_1
\right]
\Big\}.
\label{syst:Order2_x_HybridA}
\end{eqnarray}

\textbf{System 3.3.2:}
\begin{eqnarray}
\ddot{x}_1&=&\left[ 2\, x_1 \,(x_1-x_2)\right]^{-1}\Big\{2\,\dot{x}_1\,(\dot{x}_1 \,x_2 +2 \,\dot{x}_2\, x_1)
-(r_1)^2\, \omega^2 \,x_1^2\,(2 \,x_1+x_2)\notag\\
&&-\mathbf{i}\,r_3\, \omega \,x_1\,(2 \,\dot{x}_1 \,x_2+\dot{x}_2\, x_1)
\Big\}
 \notag\\
\ddot{x}_2&=&\left[ x_1\,(x_1-x_2)\right]^{-1}
\Big\{
-2\,\dot{x}_1 \,(\dot{x}_1\, x_2 +2\, \dot{x}_2 \,x_1)+(r_1)^2 \,\omega^2 \,x_1 \,x_2 \,(2\,x_1+x_2) \notag\\
&&+\mathbf{i}\, r_3\, \omega\, x_1\,(2\, \dot{x}_1 \,x_2+\dot{x}_2\, x_1)
\Big\}
\label{syst:Order2_x_HybridB}
\end{eqnarray}

\textbf{System 3.3.3:}
\begin{eqnarray}
\ddot{x}_1&=&-\left[2 \,x_1 \,(x_1-x_2) \right]^{-1}\Big\{
2\,\dot{x}_1\,\left[
\dot{x}_1\,(x_1-2 \,x_2) -2\, \dot{x}_2\, x_1
\right]
+(r_2)^2 \,\omega^2 \,(x_1)^2\,(x_1+2\,x_2)\notag \\
&& +2\, \mathbf{i}\, r_3 \,\omega \,x_1\,(2 \,\dot{x}_1\, x_2+\dot{x}_2\, x_1)
\Big\},
 \notag\\
\ddot{x}_2&=&\left[ (x_1)^2 \,(x_1-x_2) \right]^{-1} \Big\{
-2\, \dot{x}_1\, \left[\dot{x}_1\, (x_2)^2 +2 \,\dot{x}_2 \,(x_1)^2\right]+
(r_2)^2\, \omega^2 \,(x_1)^2 \,x_2\, (x_1+2 \,x_2) \notag\\
&&+\mathbf{i}\, r_3\, \omega\, x_1\,(x_1+x_2)\,(2\, \dot{x}_1 \,x_2 +\dot{x}_2\, x_1)
\Big\}.
\label{syst:Order2_x_HybridC}
\end{eqnarray}
These 3 systems are Hamiltonian,  solvable by algebraic operations and their solutions are 
\textit{isochronous}. 

Below we display the plots of the solutions of system~(\ref{syst:Order2_x_HybridC}) with the parameters
\begin{equation}
r_2=\frac{1}{3},\; \; r_3=\frac{1}{2}, \;\;\omega=2\pi,
\label{par:Order2HybridC}
\end{equation}
satisfying the initial conditions
\begin{eqnarray}
&&x_1(0) =0.94 - 0.28 \,\mathbf{i}, \;\;\dot{x}_1(0) =-0.38-4.68 \,\mathbf{i}, \notag\\
&&x_2(0) =1.40+1.11 \,\mathbf{i}, \;\;\dot{x}_2(0) = -9.20+2.50 \,\mathbf{i}.
\label{InitCond:Order2HybridC}
\end{eqnarray}

\begin{minipage}{\linewidth}
      \centering
      \begin{minipage}{0.45\linewidth}
          \begin{figure}[H]
              \includegraphics[width=\linewidth]{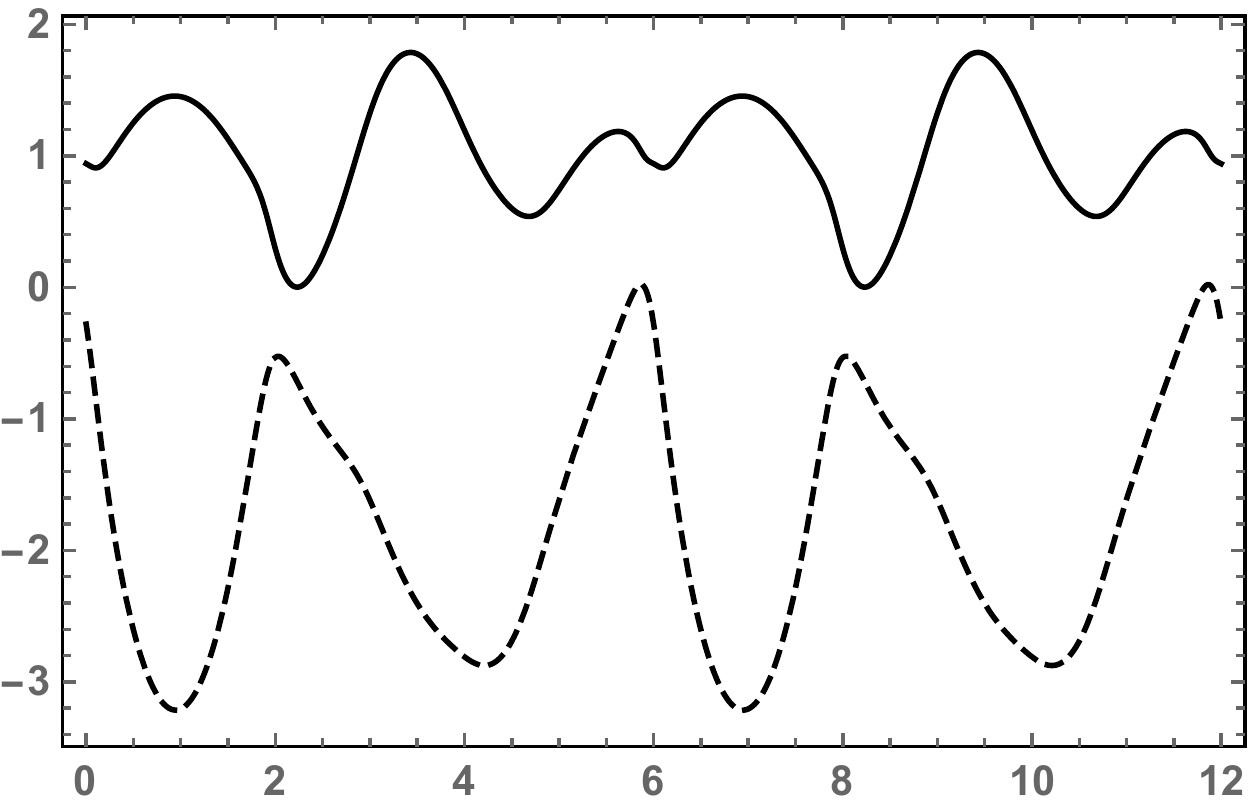}
              \caption{Initial value problem~(\ref{syst:Order2_x_HybridC}),~(\ref{par:Order2HybridC}),~(\ref{InitCond:Order2HybridC}). Graphs of the real (bold curve) and imaginary 
              (dashed curve) parts of the coordinate $x_1(t)$; period $6$.}
              \label{F1A1}
          \end{figure}
      \end{minipage}
      \hspace{0.05\linewidth}
      \begin{minipage}{0.45\linewidth}
          \begin{figure}[H]
              \includegraphics[width=\linewidth]{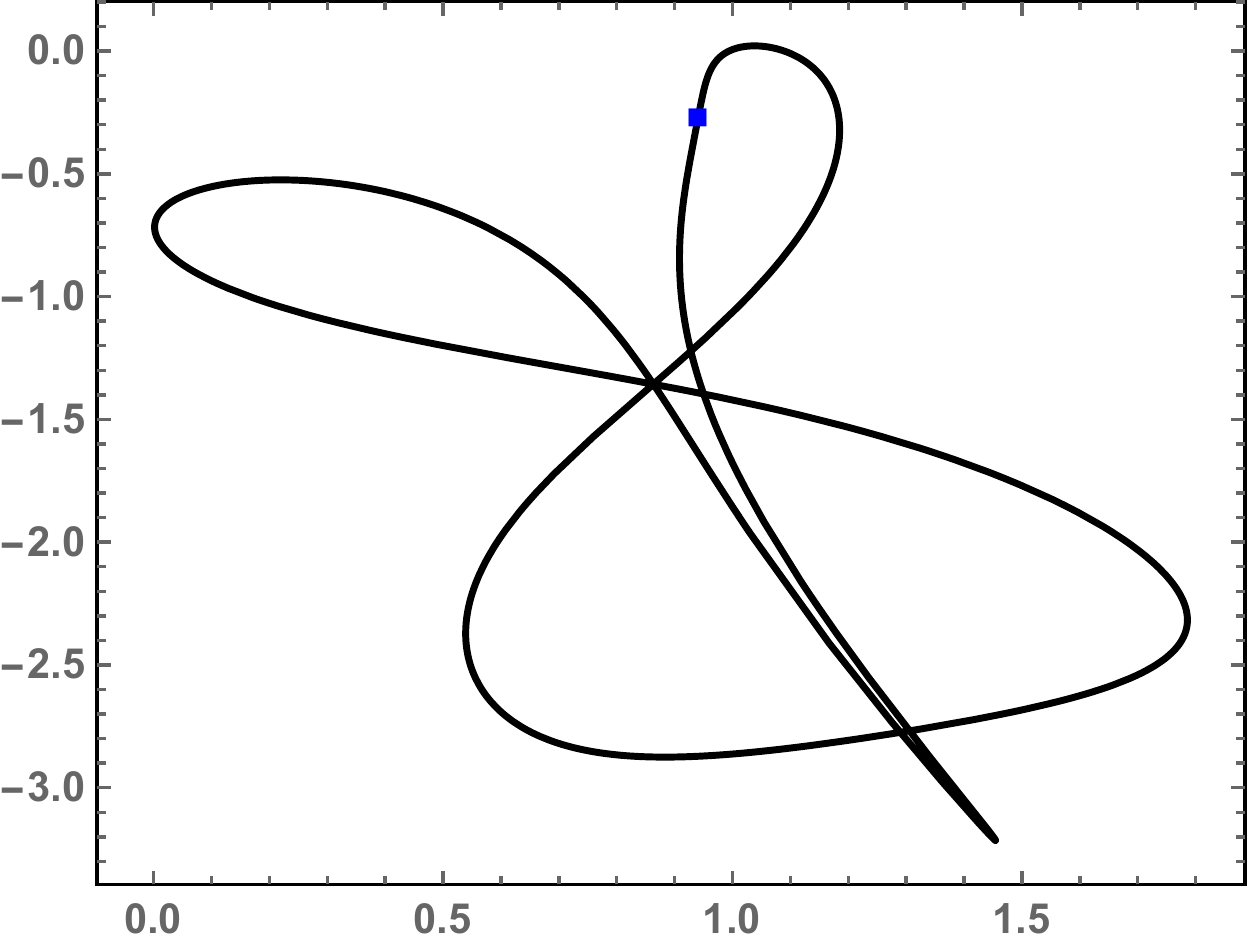}
              \caption{Initial value problem~(\ref{syst:Order2_x_HybridC}),~(\ref{par:Order2HybridC}),~(\ref{InitCond:Order2HybridC}). Trajectory, in the complex $x$-plane, of  $x_1(t)$; 
              period $6$. The   square indicates the initial condition $x_1(0)=0.94 - 0.28 \,\mathbf{i}$.}
              \label{F1A2}
          \end{figure}
      \end{minipage}
  \end{minipage}
  
  \begin{minipage}{\linewidth}
      \centering
      \begin{minipage}{0.45\linewidth}
          \begin{figure}[H]
              \includegraphics[width=\linewidth]{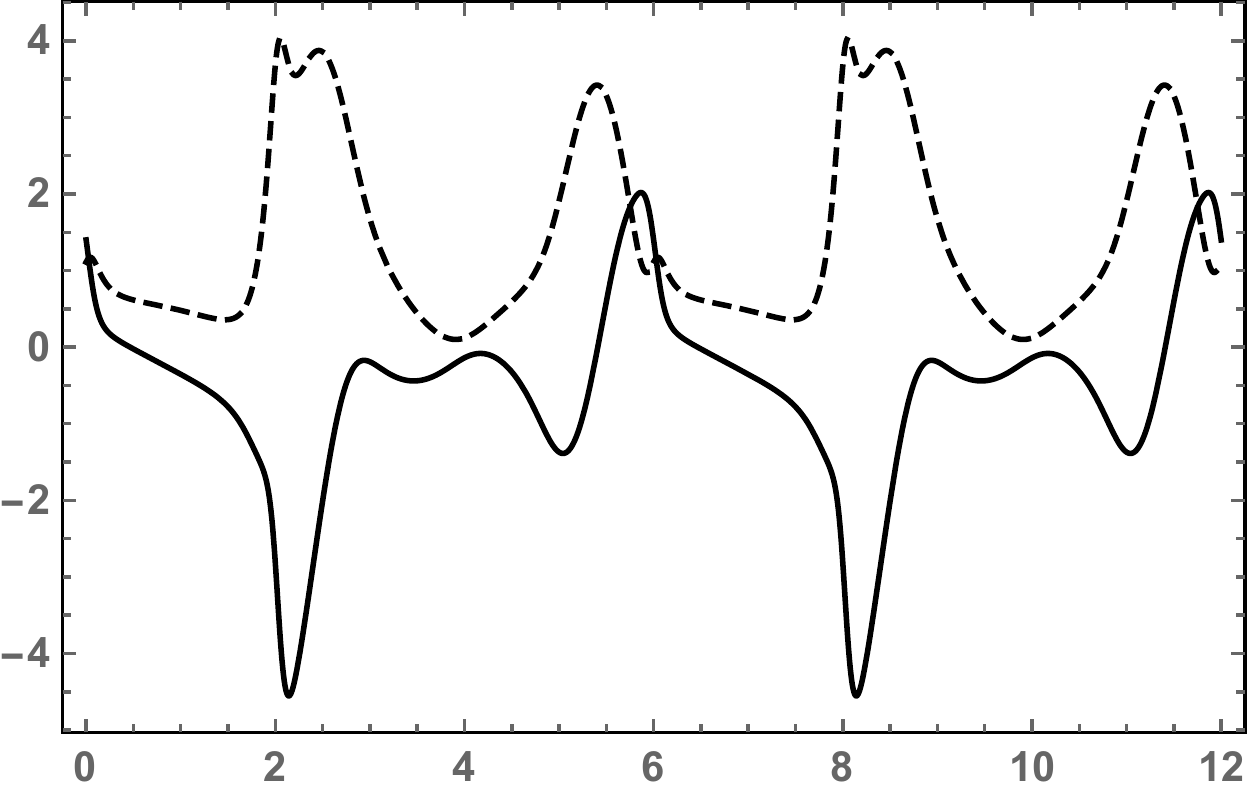}
              \caption{Initial value problem~(\ref{syst:Order2_x_HybridC}),~(\ref{par:Order2HybridC}),~(\ref{InitCond:Order2HybridC}). Graphs of the real (bold curve) and imaginary 
              (dashed curve) parts of the coordinate $x_2(t)$; period $6$.}
              \label{F1A3}
          \end{figure}
      \end{minipage}
      \hspace{0.05\linewidth}
      \begin{minipage}{0.45\linewidth}
          \begin{figure}[H]
              \includegraphics[width=\linewidth]{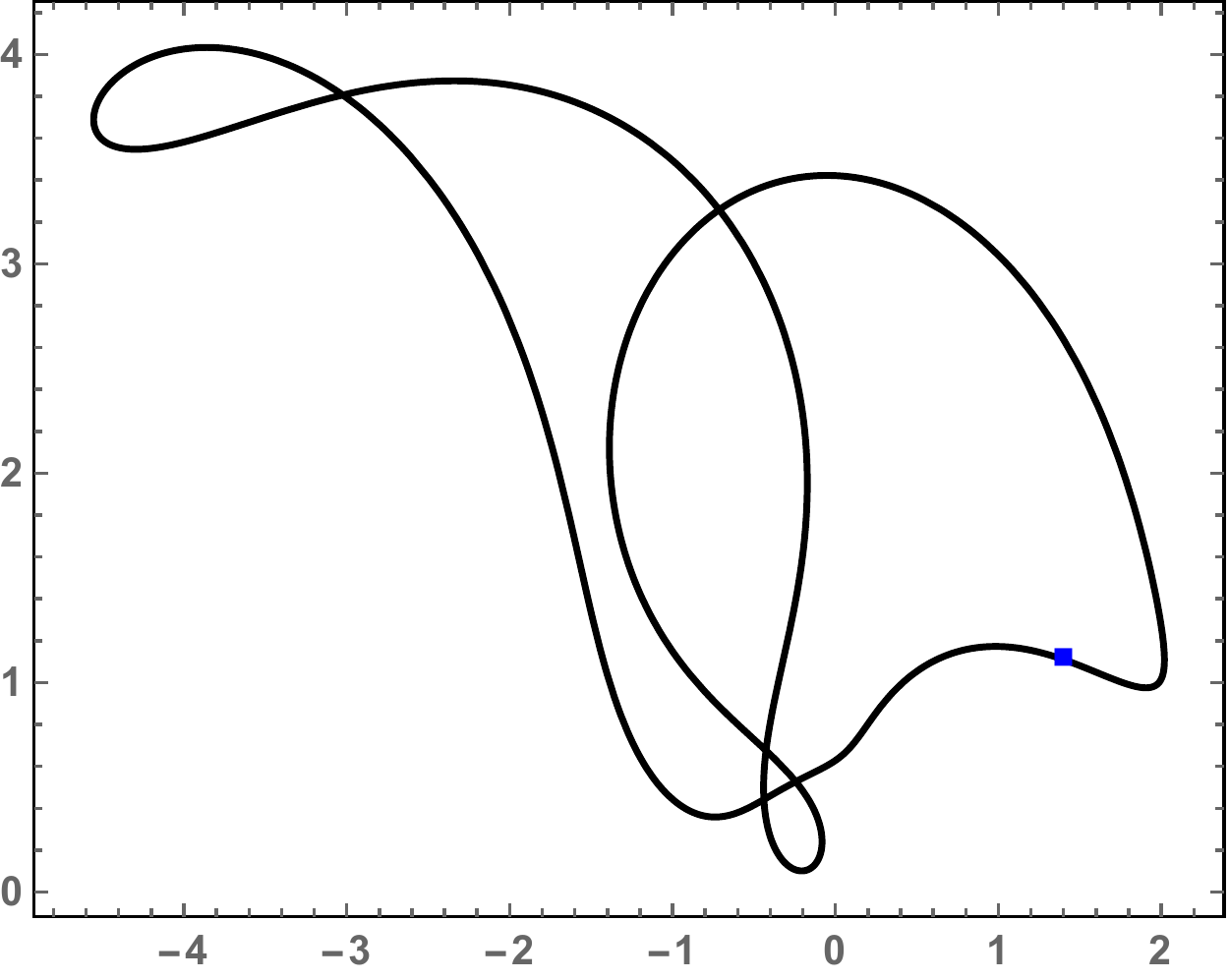}
              \caption{Initial value problem~(\ref{syst:Order2_x_HybridC}),~(\ref{par:Order2HybridC}),~(\ref{InitCond:Order2HybridC}). Trajectory, in the complex $x$-plane, of  $x_2(t)$; 
              period $6$. The   square indicates the initial condition $x_2(0)= 1.40+1.11 \,\mathbf{i}$.}
              \label{F1A4}
          \end{figure}
      \end{minipage}
  \end{minipage}

\subsection{Example 3.4}

In this example, we consider the following $3$ \textit{generating}
models: 
\begin{equation}
\text{\textbf{Model 3.4.1}: }\ddot{y}_{1}=-r^2~\omega^2 ~{y}_{1},~~~\ddot{y}%
_{2}=-a~\dot{y}_{2}~;  \label{Model31}
\end{equation}%
\begin{equation}
\text{\textbf{Model 3.4.2}: } \ddot{y}%
_{1}=-r^2~\omega^2~{y}_{1},~~~  \ddot{y}_{3}=-a~\dot{y}_{3}~;  \label{Model32}
\end{equation}%
\begin{equation}
\text{\textbf{Model 3.4.3}: }\ddot{y}_{2}=-r^2~\omega^2~{y}_{2},~~~\ddot{y}%
_{3}=-a~\dot{y}_{3}~. \label{Model33}
\end{equation}
Here $a$ is a positive real
number and $r$ is a nonvanishing rational
number. These 3 models are \textit{Hamiltonian}  (see \textbf{Remarks~3.1.1} and~\textbf{3.2.1}) and \textit{integrable}
and their solutions are given by appropriate selections from formulas~(\ref{ymModels2}) and
\begin{eqnarray}
y_m(t)=y_m(0)+\frac{1}{a} \dot{y}_m (0) \left[ 1-\exp(-a t) \right].
\label{ymModels4}
\end{eqnarray}
 For example, the solution of Model 3.4.1 is given by~(\ref{ymModels2}) with $m=1$ and~(\ref{ymModels4}) with $m=2$. 
These models are \textit{all asymptotically isochronous}~\cite{19}.

These generating models yield the following solvable two-body problems, via the method described in Subsection 2.1, see~(\ref{xdotdot1}),~(\ref{xdotdot2})  and~(\ref{xdotdot3}):

\textbf{System 3.4.1:}
\begin{eqnarray}
\ddot{x}_1&=&(x_1-x_2)^{-1}\Big\{
\dot{x}_1 \,(\dot{x}_1+2 \,\dot{x}_2)
-r^2 \,\omega^2 \,x_1\, (2 \,x_1+x_2)\notag\\
&&+a\,\left[
\dot{x}_1\,(x_1+x_2) +\dot{x}_2\, x_1
\right]
\Big\},
 \notag\\
\ddot{x}_2&=&(x_1-x_2)^{-1} \Big\{
2 \,\dot{x}_1(\dot{x}_1+ 2\,\dot{x}_2)
+r^2\, \omega^2\, (x_1+x_2)\,(2 \,x_1+x_2)\notag\\
&&-2 \,a\, \left[
\dot{x}_1\,(x_1+x_2)+\dot{x}_2\, x_1
\right]
\Big\}.
\label{syst:Order2_x_AsymptPeriodicA}
\end{eqnarray}

\textbf{System 3.4.2:}
\begin{eqnarray}
\ddot{x}_1&=&\left[2 \,x_1 \,(x_1-x_2) \right]^{-1}
\Big\{
2\, \dot{x}_1\,(\dot{x}_1\, x_2+2\,\dot{x}_2\, x_1)
-r^2\, \omega^2\, (x_1)^2\,(2 \,x_1+x_2)\notag\\
&&+ a\, x_1\,(2\, \dot{x}_1 \,x_2+\dot{x}_2\, x_1)
\Big\},
 \notag\\
\ddot{x}_2&=&\left[ x_1\,(x_1-x_2) \right]^{-1}
\Big\{
-2 \,\dot{x}_1\, (\dot{x}_1 \,x_2 +2\, \dot{x}_2\, x_1)+
r \,\omega^2\, x_1 \,x_2 (2\, x_1+ x_2) \notag\\
&&- a \,x_1\, (2 \,\dot{x}_1 \,x_2 +\dot{x}_2\, x_1)
\Big\}.
\label{syst:Order2_x_AsymptPeriodicB}
\end{eqnarray}

\textbf{System 3.4.3:}
\begin{eqnarray}
\ddot{x}_1&=&\left[ 2 \,x_1 \,(x_1-x_2) \right]^{-1}
\Big\{
- 2 \,\dot{x}_1 \,\left[
\dot{x}_1 \,(x_1-2\, x_2) -2\, \dot{x}_2\, x_1
\right]\notag\\
&&
-r^2 \,\omega^2 \,(x_1)^2 \,(x_1+2\, x_2) 
+ 2 \,a\, x_1\, (2 \,\dot{x}_1\, x_2+\dot{x}_2\, x_1)
\Big\},
 \notag\\
\ddot{x}_2&=&\left[ (x_1)^2\, (x_1-x_2) \right]^{-1}
\Big\{
-2\, \dot{x}_1\, \left[\dot{x}_1 \,(x_2)^2+2 \,\dot{x}_2 (x_1)^2\right]\notag\\
&&+r^2 \,\omega^2 \,(x_1)^2\, x_2\, (x_1+2\, x_2)
-a \,x_1\,(x_1+x_2) (2\, \dot{x}_1\, x_2+\dot{x}_2\, x_1)
\Big\}.
\label{syst:Order2_x_AsymptPeriodicC}
\end{eqnarray}
These 3 systems are Hamiltonian,  solvable by algebraic operations and their solutions are 
\textit{asymptotically isochronous}. 

Below we display the plots of the solutions of system~\eqref{syst:Order2_x_AsymptPeriodicB} with the parameters
\begin{equation}
r=\frac{1}{3}, \;\;\omega=2\pi, \;\;a=0.1,
\label{par:Order2AsymptPeriodicB}
\end{equation}
satisfying the initial conditions
\begin{eqnarray}
&&x_1(0) =1.21, \;\;\dot{x}_1(0) =-0.56-2.34 \,\mathbf{i}, \notag\\
&&x_2(0) =1.42+0.89 \,\mathbf{i}, \;\;\dot{x}_2(0) =-1.78-0.54\,\mathbf{i}.
\label{InitCond:Order2AsymptPeriodicB}
\end{eqnarray}

\begin{minipage}{\linewidth}
      \centering
      \begin{minipage}{0.45\linewidth}
          \begin{figure}[H]
              \includegraphics[width=\linewidth]{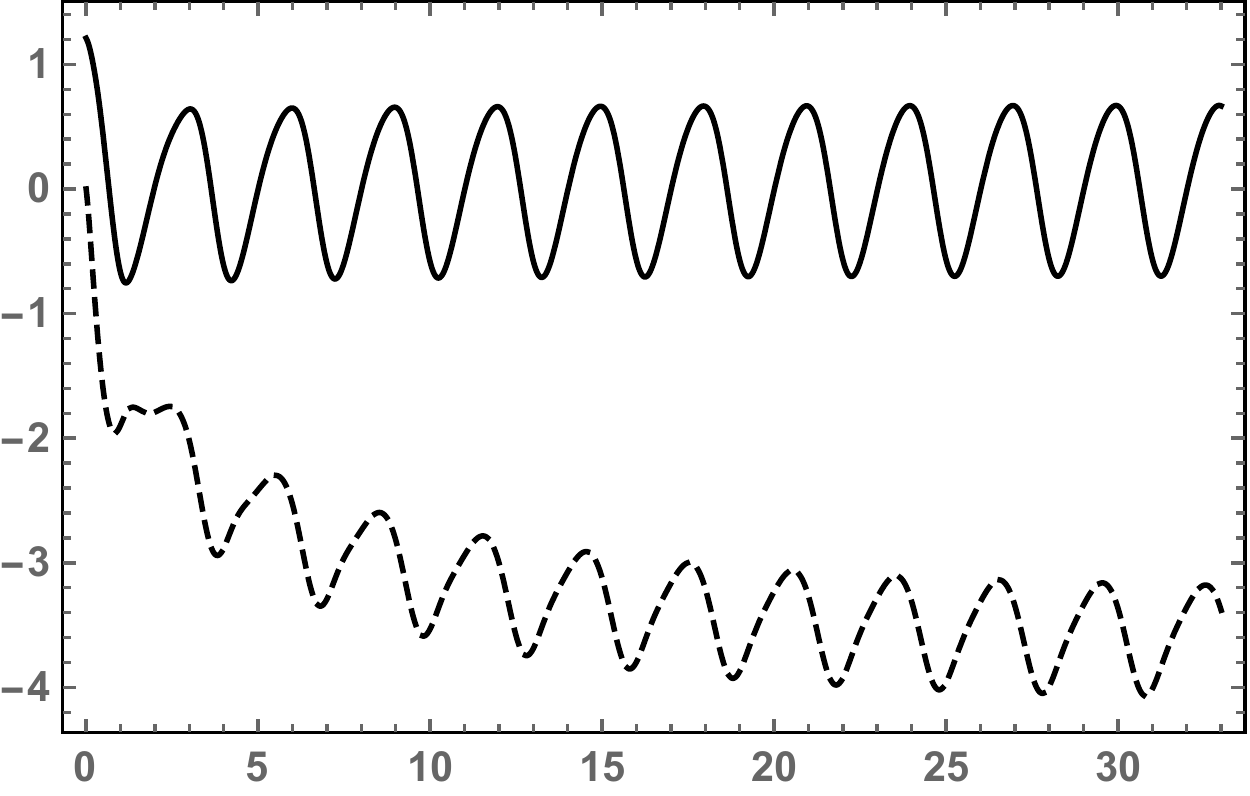}
              \caption{Initial value problem~(\ref{syst:Order2_x_AsymptPeriodicB}),~(\ref{par:Order2AsymptPeriodicB}),~(\ref{InitCond:Order2AsymptPeriodicB}). Graphs of the real (bold curve) and imaginary 
              (dashed curve) parts of the coordinate $x_1(t)$.}
              \label{F1A1}
          \end{figure}
      \end{minipage}
      \hspace{0.05\linewidth}
      \begin{minipage}{0.45\linewidth}
          \begin{figure}[H]
              \includegraphics[width=\linewidth]{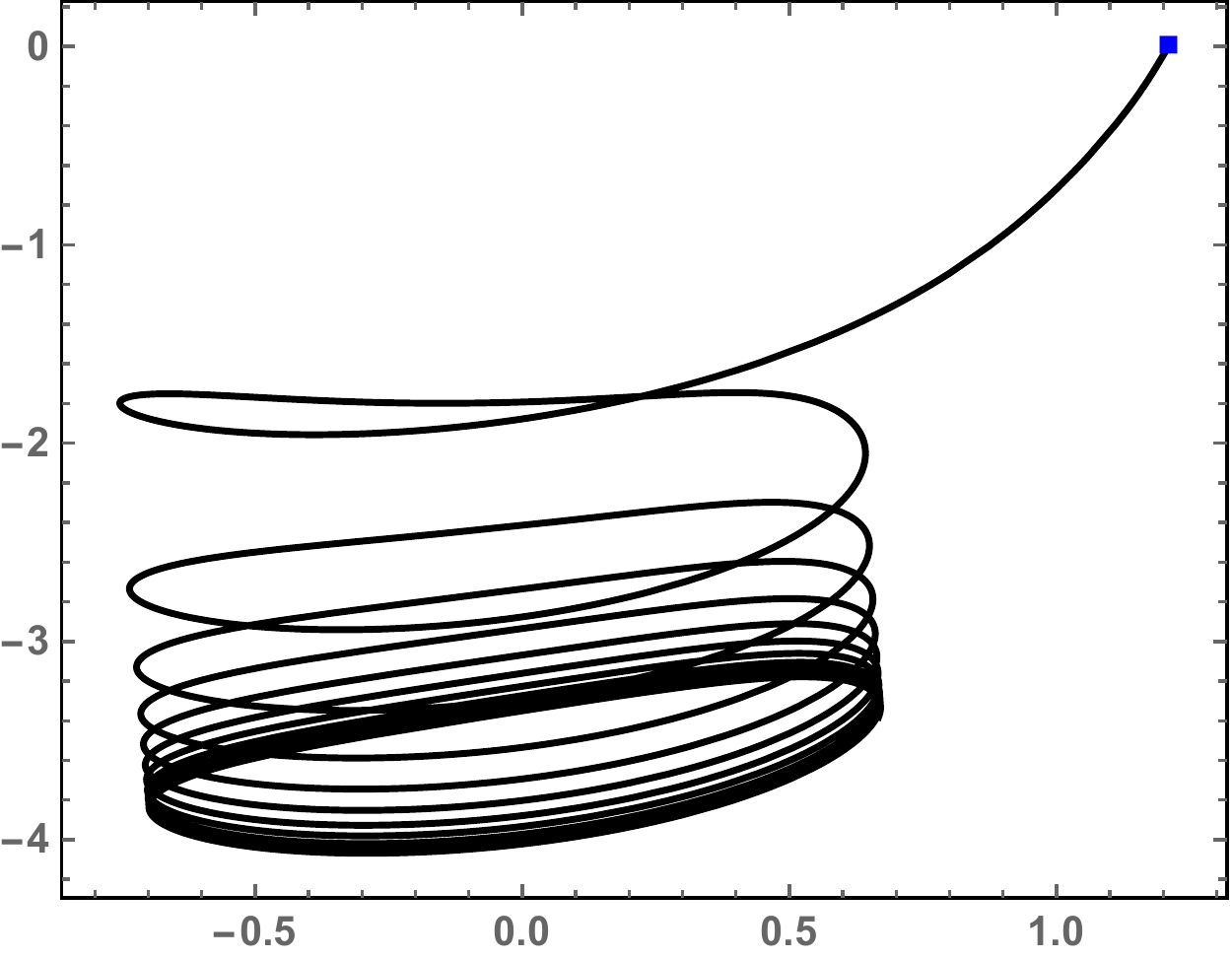}
              \caption{Initial value problem~(\ref{syst:Order2_x_AsymptPeriodicB}),~(\ref{par:Order2AsymptPeriodicB}),~(\ref{InitCond:Order2AsymptPeriodicB}). Trajectory, in the complex $x$-plane, of  $x_1(t)$. The   square indicates the initial condition $x_1(0)=1.21$.}
              \label{F1A2}
          \end{figure}
      \end{minipage}
  \end{minipage}
  
  \begin{minipage}{\linewidth}
      \centering
      \begin{minipage}{0.45\linewidth}
          \begin{figure}[H]
              \includegraphics[width=\linewidth]{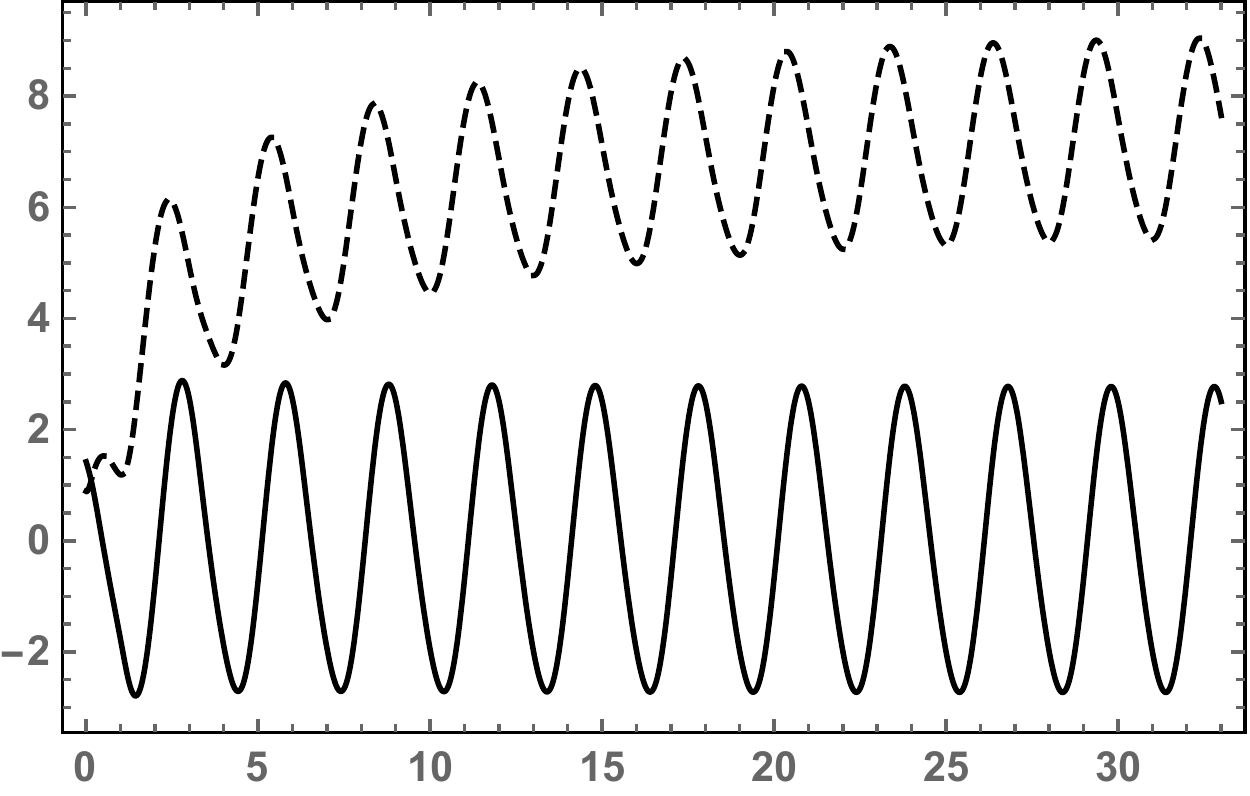}
              \caption{Initial value problem~(\ref{syst:Order2_x_AsymptPeriodicB}),~(\ref{par:Order2AsymptPeriodicB}),~(\ref{InitCond:Order2AsymptPeriodicB}). Graphs of the real (bold curve) and imaginary 
              (dashed curve) parts of the coordinate $x_2(t)$.}
              \label{F1A3}
          \end{figure}
      \end{minipage}
      \hspace{0.05\linewidth}
      \begin{minipage}{0.45\linewidth}
          \begin{figure}[H]
              \includegraphics[width=\linewidth]{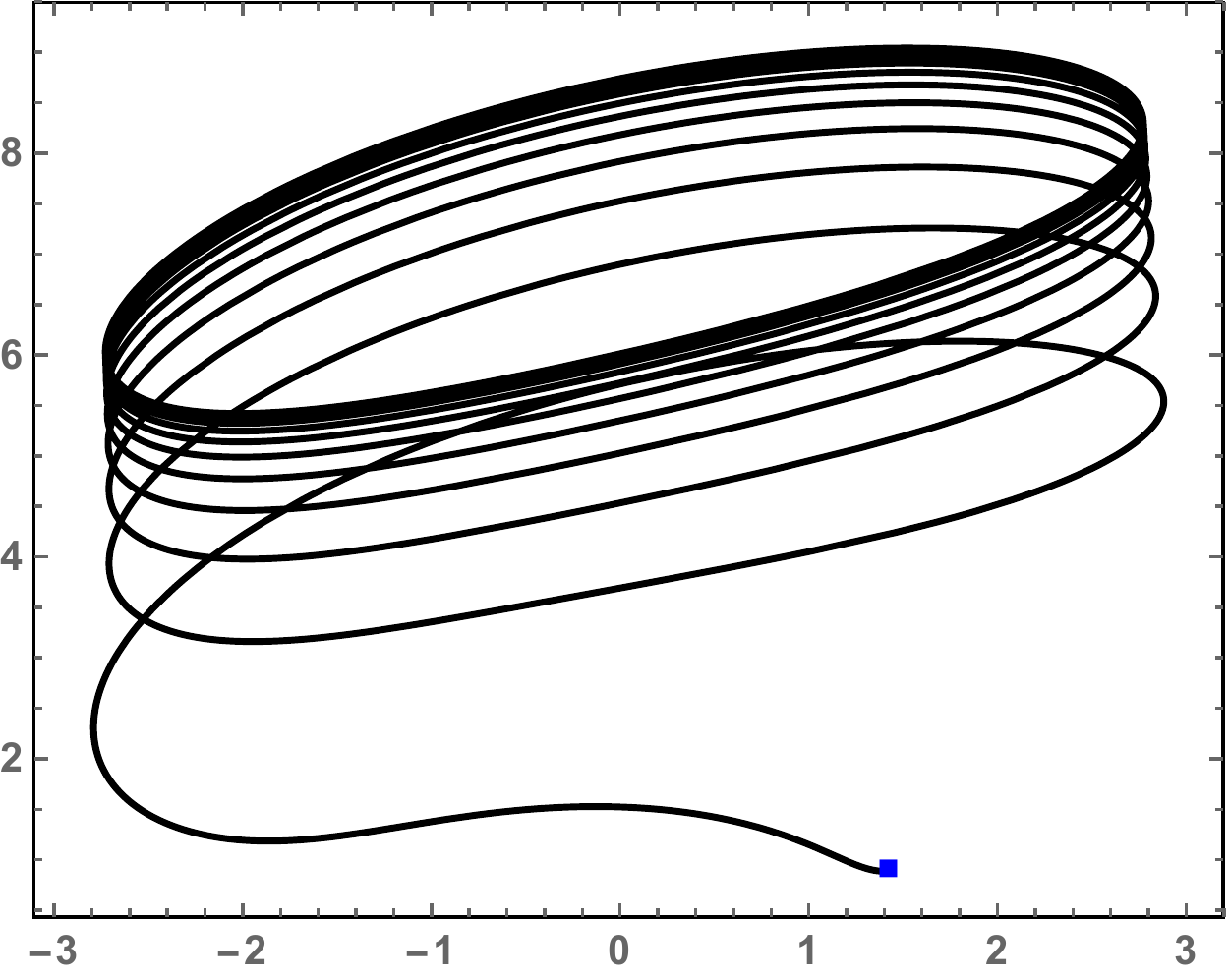}
              \caption{Initial value problem~(\ref{syst:Order2_x_AsymptPeriodicB}),~(\ref{par:Order2AsymptPeriodicB}),~(\ref{InitCond:Order2AsymptPeriodicB}). Trajectory, in the complex $x$-plane, of  $x_2(t)$. The   square indicates the initial condition $x_2(0)=1.42+0.89 \,\mathbf{i}$.}
              \label{F1A4}
          \end{figure}
      \end{minipage}
  \end{minipage}

\subsection{Example 3.5}

In this example, we take as a starting point of our treatment one of the following $4$ \textit{generating}
models: 
\begin{eqnarray}
\text{\textbf{Model 3.5.1}: }\ddot{y}_{1}=- r_1^2~\omega^2~{y}_{1},~~~\ddot{y}%
_{2}=- r_2^2 ~\omega^2~{y}_{2}, 
~~~\ddot{y}%
_{3}=- r_3^2 ~\omega^2~{y}_{3}~;  \label{Model51}
\end{eqnarray}%
\begin{eqnarray}
\text{\textbf{Model 3.5.2}: } \ddot{y}%
_{1}=- r_1^2~ \omega^2~{y}_{1},~~~  \ddot{y}_{2}=- r_2^2~ \omega^2~{y}_{2}, 
~~~\ddot{y}%
_{4}=- r_4^2~ \omega^2~{y}_{4}~;
\label{Model52}
\end{eqnarray}%
\begin{eqnarray}
\text{\textbf{Model 3.5.3}: }\ddot{y}_{1}=- r_1^2~ \omega^2~{y}_{1},~~~\ddot{y}%
_{3}= - r_3^2~ \omega^2~{y}_{3}, 
~~~\ddot{y}%
_{4}= - r_4^2~ \omega^2~{y}_{4}~;
\label{Model53}
\end{eqnarray}
\begin{eqnarray}
\text{\textbf{Model 3.5.4}: } \ddot{y}_{2}=- r_2^2~ \omega^2~{y}_{2},~~~\ddot{y}%
_{3}=- r_3^2~ \omega^2~{y}_{3}, 
~~~\ddot{y}%
_{4}=- r_4^2~ \omega^2~{y}_{4}~.
\label{Model54}
\end{eqnarray}

Similarly to Example~1, $\omega $ is an arbitrary nonvanishing real
number and $r_{1},~r_{2},~r_{3},~r_4$ are $4$ arbitrary nonvanishing rational
numbers. These 4 models are \textit{Hamiltonian} (see \textbf{Remark 3.2.1}) and \textit{integrable}
and their solutions (see (\ref{ymModels2}) with $m=1,2,3,4$)
are \textit{isochronous} with a period which is an \textit{integer multiple}
of the basic period~(\ref{T}).

These generating models yield the following \textit{four} solvable three-body problems, via the method described in \textbf{Subsection 2.2}, see~(\ref{x1ndotdot}). More precisely, each  Model~5.(${5-\bar{m}}$)  generates 

\textbf{System 3.5.($\mathbf{5-\bar{m}}$), where $\mathbf{\bar{m}\in\{1,2,3,4\}}$:}
\begin{subequations}
\begin{eqnarray}
\ddot{x}_1&=&-(4-\bar{m}) \frac{(\dot{x}_1)^2}{x_1}+\frac{\dot{x}_1(2 \,\dot{x}_2+\dot{x}_1)}{x_1-x_2}
+\frac{\dot{x}_1(2 \,\dot{x}_3+\dot{x}_1)}{x_1-x_3}\notag\\
&&-\frac{1}{2\,(x_1-x_2)\,(x_1-x_3)} \sum_{m=1, m\neq \bar{m}}^4 \,(m-\bar{m}) \,(r_m)^2 \,\omega^2 (\,x_1)^{3-m} \,y_m,\notag\\
\ddot{x}_2&=&\frac{2 \,\dot{x}_2 \,\dot{x}_3}{x_2-x_3} +\frac{2 \,\dot{x}_1}{x_2-x_1}\,\Big[
2\,\dot{x}_2+\left(\frac{x_2}{x_1}\right)^{4-\bar{m}} \,\left(\frac{x_1-x_3}{x_2-x_3} \right) \,\dot{x}_1
\Big]\notag\\
&&+\frac{(x_2)^{4-\bar{m}}}{(x_2-x_1)\,(x_2-x_3)} \sum_{m=1, m\neq \bar{m}}^4 \,(r_m)^2\, \omega^2 \,y_m \left[ \frac{(x_2)^{\bar{m}-m}-(x_1)^{\bar{m}-m}}{x_2-x_1} \right], \notag\\
\ddot{x}_3&=&-\frac{2\, \dot{x}_2 \,\dot{x}_3}{x_2-x_3} +\frac{2\, \dot{x}_1}{x_3-x_1}\Big[
2\,\dot{x}_3+\left(\frac{x_3}{x_1}\right)^{4-\bar{m}} \left(\frac{x_1-x_2}{x_3-x_2} \right) \dot{x}_1
\Big]\notag\\
&&+\frac{(x_3)^{4-\bar{m}}}{(x_3-x_1)\,(x_3-x_2)} \sum_{m=1, m\neq \bar{m}}^4\, (r_m)^2\, \omega^2\, y_m \left[ \frac{(x_3)^{\bar{m}-m}-(x_1)^{\bar{m}-m}}{x_3-x_1} \right], \notag\\
\end{eqnarray}
where
\begin{eqnarray}
&&y_1=-(2\,x_1 +x_2+x_3),\notag\\
&&y_2=(x_1)^2+2\, x_1 \,x_2+2\, x_1 \,x_3 + x_2 \,x_3, \notag\\
&&y_3=-\left[ (x_1)^2 \,x_2+(x_1)^2\, x_3 +2 \,x_1\, x_2 \,x_3 \right],\notag\\
&&y_4= (x_1)^2\, x_2 \,x_3,
\end{eqnarray}
\label{syst:Order2_x_3Body}
\end{subequations}
see~(\ref{ymxSect22}).
These 4 systems are Hamiltonian,  solvable by algebraic operations and their solutions are 
\textit{isochronous}.

Below we display the plots of the solutions of system~\eqref{syst:Order2_x_3Body}  for the case 
\begin{equation}
\bar{m}=3
\label{mbarValue}
\end{equation}
 with the parameters
\begin{equation}
r_1=\frac{1}{2}, ~r_2=\frac{1}{3},~ r_4=\frac{2}{3}; \;\;\omega=2\pi, 
\label{par:Order2_x_3Body}
\end{equation}
satisfying the initial conditions
\begin{eqnarray}
&&x_1(0) = 1.74 + 1.42\,\mathbf{i}, \;\;\dot{x}_1(0) =12.47 + 4.46 \,\mathbf{i}, \notag\\
&&x_2(0) =-3.20 + 0.52 \,\mathbf{i}, \;\;\dot{x}_2(0) =-10.23 + 6.40 \,\mathbf{i}, \notag\\
&&x_3(0) =0.44 - 3.15 \,\mathbf{i}, \;\;\dot{x}_3(0) =3.16 - 14.66 \,\mathbf{i}.
\label{InitCond:Order2_x_3Body}
\end{eqnarray}

\begin{minipage}{\linewidth}
      \centering
      \begin{minipage}{0.45\linewidth}
          \begin{figure}[H]
              \includegraphics[width=\linewidth]{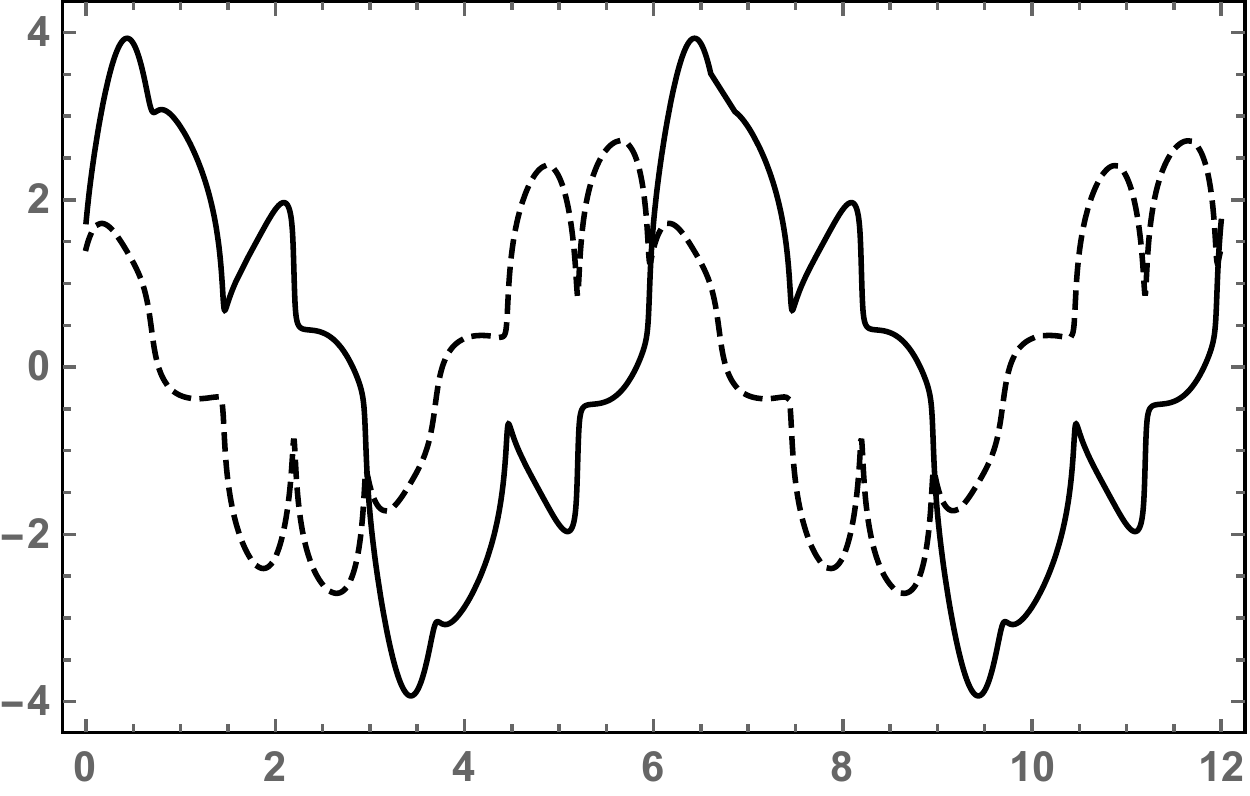}
              \caption{Initial value problem~(\ref{syst:Order2_x_3Body}),~(\ref{mbarValue}),~(\ref{par:Order2_x_3Body}),~(\ref{InitCond:Order2_x_3Body}). Graphs of the real (bold curve) and imaginary 
              (dashed curve) parts of the coordinate $x_1(t)$; period 6.}
              \label{F1A1}
          \end{figure}
      \end{minipage}
      \hspace{0.05\linewidth}
      \begin{minipage}{0.45\linewidth}
          \begin{figure}[H]
              \includegraphics[width=\linewidth]{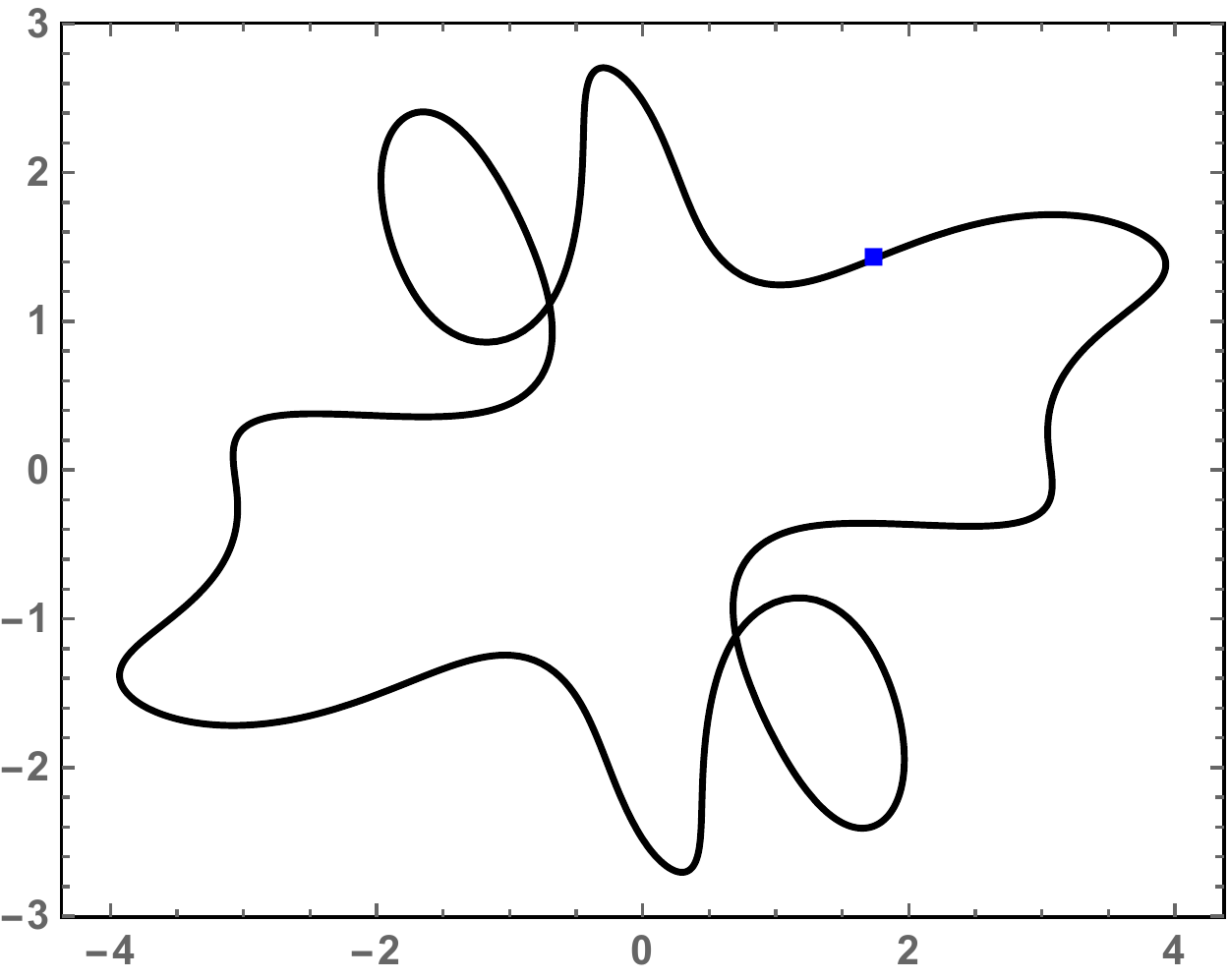}
              \caption{Initial value problem~(\ref{syst:Order2_x_3Body}),~(\ref{mbarValue}),~(\ref{par:Order2_x_3Body}),~(\ref{InitCond:Order2_x_3Body}). Trajectory, in the complex $x$-plane, of  $x_1(t)$. The   square indicates the initial condition $x_1(0)= 1.74 + 1.42\,\mathbf{i}$.}
              \label{F1A2}
          \end{figure}
      \end{minipage}
  \end{minipage}
  
\begin{minipage}{\linewidth}
      \centering
      \begin{minipage}{0.45\linewidth}
          \begin{figure}[H]
              \includegraphics[width=\linewidth]{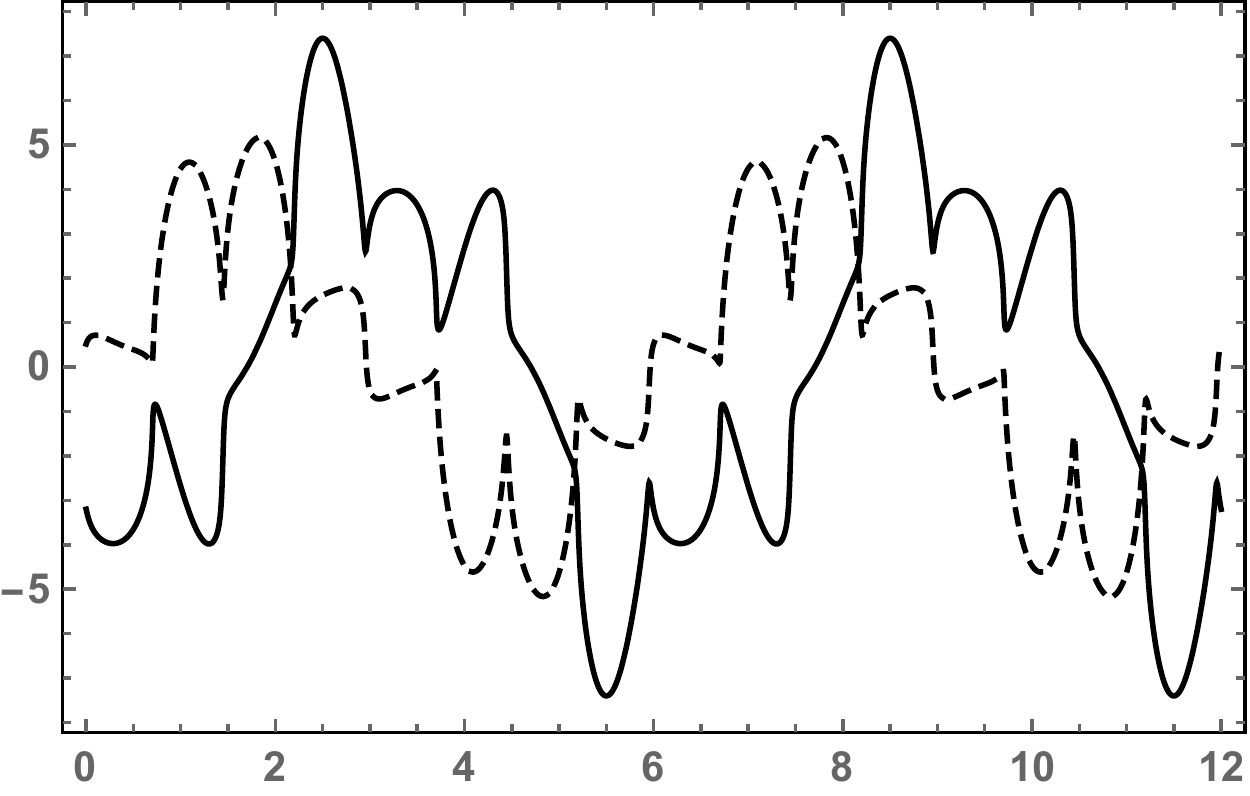}
              \caption{Initial value problem~(\ref{syst:Order2_x_3Body}),~(\ref{mbarValue}),~(\ref{par:Order2_x_3Body}),~(\ref{InitCond:Order2_x_3Body}). Graphs of the real (bold curve) and imaginary 
              (dashed curve) parts of the coordinate $x_2(t)$; period 6.}
              \label{F1A1}
          \end{figure}
      \end{minipage}
      \hspace{0.05\linewidth}
      \begin{minipage}{0.45\linewidth}
          \begin{figure}[H]
              \includegraphics[width=\linewidth]{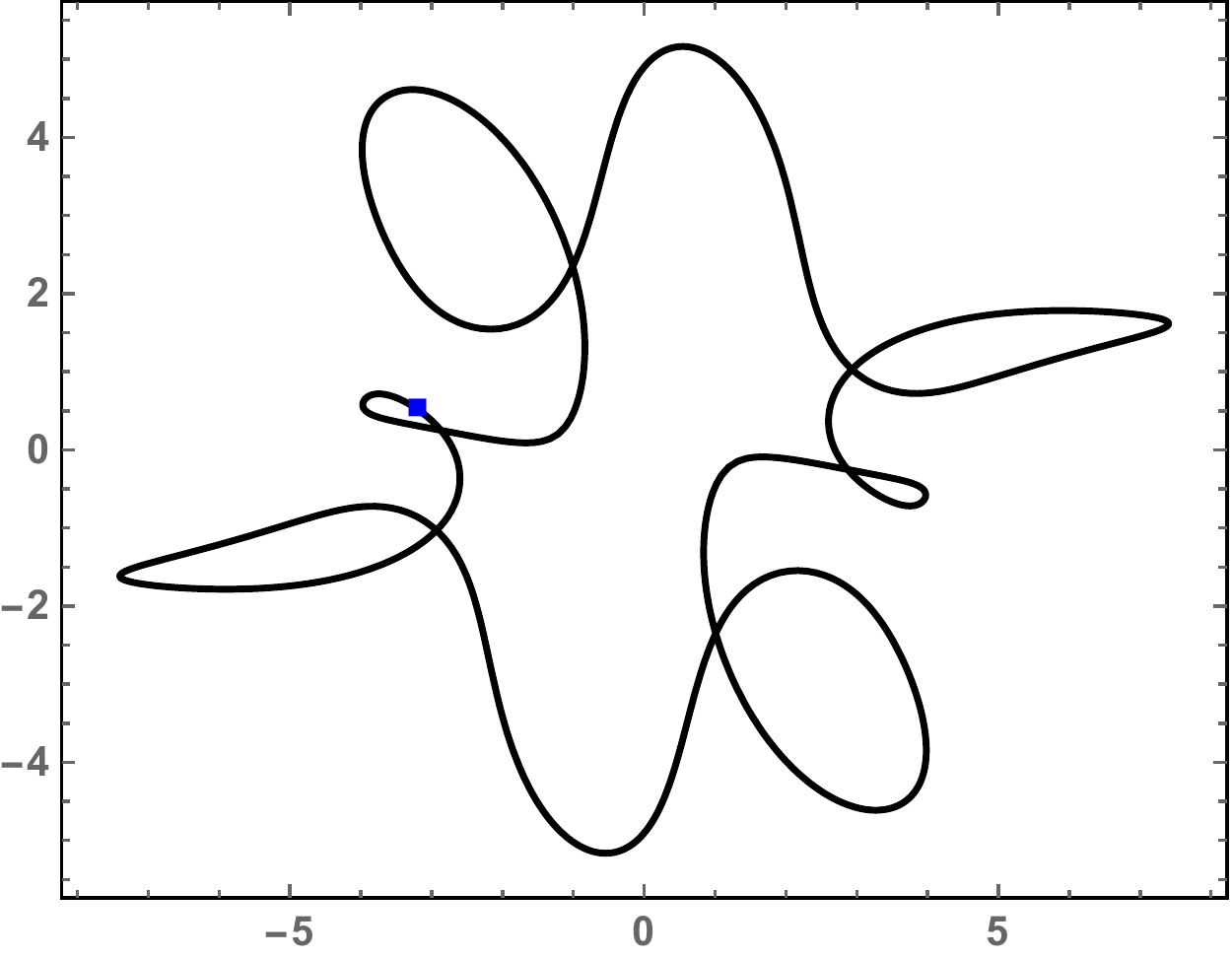}
              \caption{Initial value problem~(\ref{syst:Order2_x_3Body}),~(\ref{mbarValue}),~(\ref{par:Order2_x_3Body}),~(\ref{InitCond:Order2_x_3Body}). Trajectory, in the complex $x$-plane, of  $x_2(t)$. The   square indicates the initial condition $x_2(0)= -3.20 + 0.52 \,\mathbf{i}$.}
              \label{F1A2}
          \end{figure}
      \end{minipage}
  \end{minipage}
  
  \begin{minipage}{\linewidth}
      \centering
      \begin{minipage}{0.45\linewidth}
          \begin{figure}[H]
              \includegraphics[width=\linewidth]{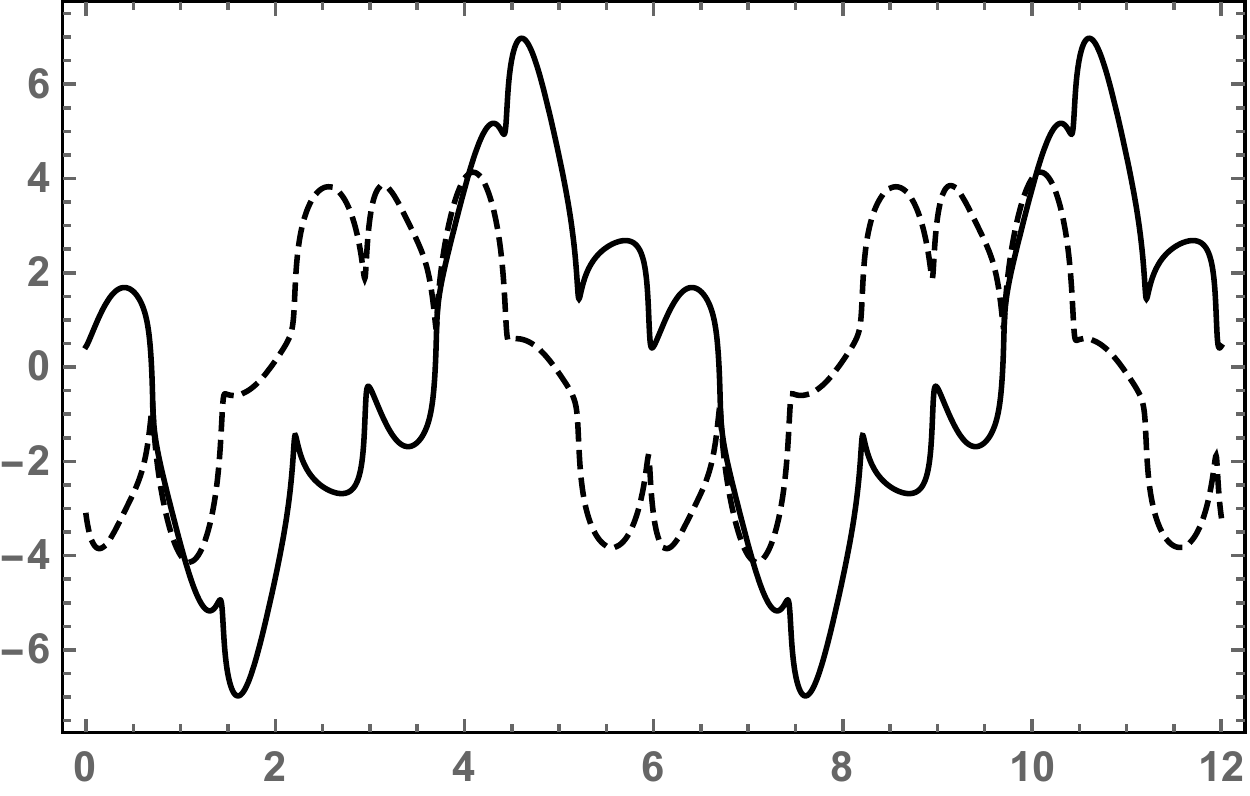}
              \caption{Initial value problem~~(\ref{syst:Order2_x_3Body}),~(\ref{mbarValue}),~(\ref{par:Order2_x_3Body}),~(\ref{InitCond:Order2_x_3Body}). Graphs of the real (bold curve) and imaginary 
              (dashed curve) parts of the coordinate $x_3(t)$; period 6.}
              \label{F1A1}
          \end{figure}
      \end{minipage}
      \hspace{0.05\linewidth}
      \begin{minipage}{0.45\linewidth}
          \begin{figure}[H]
              \includegraphics[width=\linewidth]{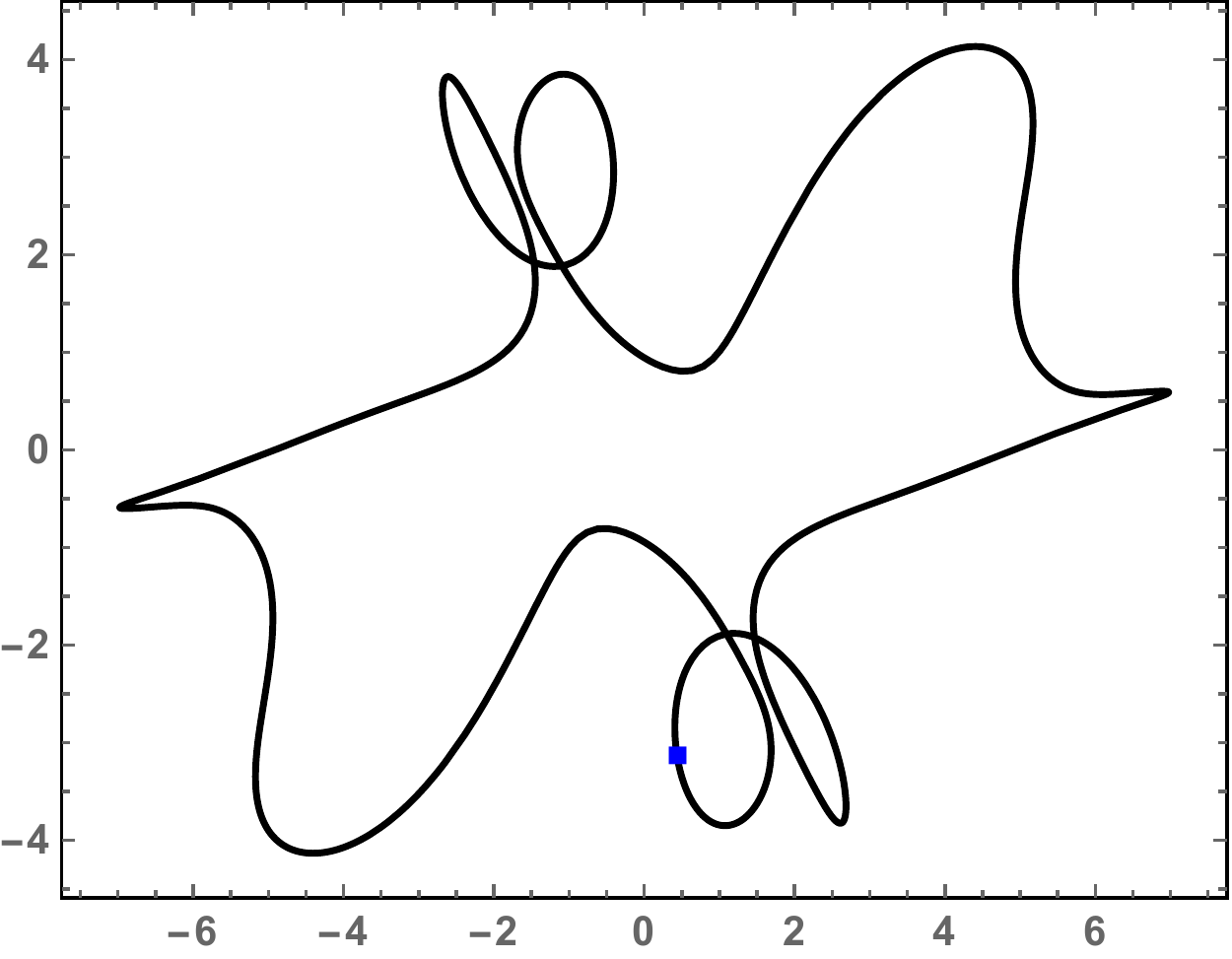}
              \caption{Initial value problem~(\ref{syst:Order2_x_3Body}),~(\ref{mbarValue}),~(\ref{par:Order2_x_3Body}),~(\ref{InitCond:Order2_x_3Body}). Trajectory, in the complex $x$-plane, of  $x_3(t)$. The   square indicates the initial condition $x_3(0)= 0.44 - 3.15 \,\mathbf{i}$.}
              \label{F1A2}
          \end{figure}
      \end{minipage}
  \end{minipage}

\section{Outlook}

In this Section 4 we tersely outline further developments which are a
natural continuation of the findings reported in this paper.

Two kinds of generalizations of the results reported in this paper are
obvious goals. One---already mentioned at the end of the introductory part
of Section 2---is to extend the results of this paper---which are confined
to time-dependent polynomials of \textit{arbitrary} degree in the complex
variable $z$ featuring, for \textit{all} time, a \textit{one double} 
zero---to the most general case of analogous polynomials featuring, for 
\textit{all} time, \textit{several} zeros, each with an \textit{%
arbitrary} (fixed) multiplicity. Another direction of generalization is to obtain
formulas---say, analogous to (13) and (30)---expressing time-derivatives 
\textit{of order} $k>2$ of the zeros $x_{n}\left( t\right) $ of such
time-dependent polynomials.

And an unlimited area of additional study is of course open, consisting in
the identification and investigation of \textit{new} many-body problems in
the plane amenable to \textit{exact} treatments via techniques analogous to
those demonstrated by the few examples treated above (see Section 3) and in
previous publications \cite{1}-\cite{14}---including possible applications of these
findings.

\end{document}